\documentclass[aip,jcp,floatfix,reprint]{revtex4-1}

\usepackage[T1]{fontenc}       

\usepackage{ae,aecompl}
\usepackage{graphicx}
\usepackage{epstopdf}
\usepackage{color}
\usepackage{amsmath}
\usepackage{hyperref}
\usepackage{amsfonts}

\newcommand{\comment}[1]{}

\global\long\def\M#1{\boldsymbol{#1}}

\global\long\def\grad{\M{\nabla}}

\begin{document}

\title{Rapid Sampling of Stochastic Displacements in Brownian Dynamics Simulations with Stresslet Constraints}

\author{Andrew M. Fiore}
\author{James W. Swan}
\email{jswan@mit.edu}
\affiliation{Department of Chemical Engineering, Massachusetts Institute of Technology, Cambridge, MA 02139, USA}

\keywords{Colloids, Brownian Motion, Hydrodynamics, Coarse Graining, Constraints, Simulations}

\begin{abstract}

Brownian Dynamics simulations are an important tool for modeling the dynamics of soft matter.  However, accurate and rapid computations of the hydrodynamic interactions between suspended, microscopic components in a soft material is a significant computational challenge.  Here, we present a new method for Brownian Dynamics simulations of suspended colloidal scale particles such as colloids, polymers, surfactants, and proteins subject to a particular and important class of hydrodynamic constraints. The total computational cost of the algorithm is practically linear with the number of particles modeled and can be further optimized when the characteristic mass fractal dimension of the suspended particles is known. Specifically, we consider the so-called ``stresslet'' constraint for which suspended particles resist local deformation.  This acts to produce a symmetric force dipole in the fluid and imparts rigidity to the particles.
The presented method is an extension of the recently reported positively-split formulation for Ewald summation of the Rotne-Prager-Yamakawa (RPY) mobility tensor to higher order terms in the hydrodynamic scattering series accounting for force dipoles [Andrew M. Fiore, Florencio Balboa Usabiaga, Aleksandar Donev, and James W. Swan. Rapid sampling of stochastic displacements in brownian dynamics simulations. The Journal of Chemical Physics, 146(12):124116, 2017]. The hydrodynamic mobility tensor, which is proportional to the covariance of particle Brownian displacements, is constructed as an Ewald sum in a novel way which guarantees that the real-space and wave-space contributions to the sum are independently symmetric and positive-definite for all possible particle configurations.  This property of the Ewald sum is leveraged to rapidly sample the Brownian displacements from a superposition of statistically independent processes with the wave-space and real-space contributions as respective covariances. The cost of computing the Brownian displacements in this way is comparable to the cost of computing the deterministic displacements.
Addition of a stresslet constraint to the over-damped particle equations of motion leads to a stochastic differential algebraic equation (SDAE) of index 1, which is integrated forward in time using a mid-point integration scheme that implicitly produces stochastic displacements consistent with the fluctuation-dissipation theorem for the constrained system.
Calculations for hard sphere dispersions are illustrated and used to explore the performance of the algorithm. An open source, high-performance implementation on graphics processing units capable of dynamic simulations of millions of particles and integrated with the software package HOOMD-blue is used for benchmarking and made freely available in the supplementary material. 

\end{abstract}

\maketitle

\section{Introduction}
\label{sec:intro}

	The dynamics of soft matter systems are often controlled by long-range fluid-mediated interactions, called hydrodynamic interactions (HI), that couple the motion of constituent elements. As particles (colloidal particles, polymers, proteins, cells, etc.) move through a fluid they induce a velocity field that entrains other particles in the suspension. The coupling of particle motion by HI is fundamental to the dynamics of a wide range of processes. 
	HI are responsible for controlling the relative rates of cluster growth and aggregation in colloidal gelation and recent work has shown that they are crucial in determining the phase boundary in kinetically arrested systems\cite{zsigi}, as well as in setting the rate of microstructural evolution over both short and long time scales\cite{spinodal}. 
	The transport properties of polymeric materials such as DNA and proteins cannot be accurately predicted by models that neglect HI\cite{shaqfeh,protein}. Therefore, HI are necessary to describe the rheology of dilute and semi-dilute polymer solutions, and for understanding biological processes such as protein folding. 
	Gravitational settling of colloidal particles is another process driven by HI, with applications in industrial separations such as slurry dewatering and ultracentrifugation, among many others\cite{sedimentation1,sedimentation2}. 
	In all cases, fluid flow between and around particles controls the rates of transport processes and couples the motion of even well-separated particles. 
	
	 In a Newtonian fluid, HIs produce a many-body coupling between the motion of suspended micro-particles that is dependent on the particle configuration and determined by the solution of the Stokes equations subject to constraints and boundary conditions on the surfaces of all the immersed particles. Because the many-body problem is difficult to solve, various approximations for the solution to the Stokes equations have been developed and employed in the simulation of soft materials. Explicit solvent techniques such as Lattice-Boltzmann (LB)\cite{lb}, Dissipative Particle Dynamics (DPD)\cite{dpd}, and Multi-particle Collision Dynamics (MPC)\cite{mpc} do not solve the continuum over-damped momentum conservation (Stokes) equations directly, but rather represent the fluid motion explicitly as a collection of momentum-conserving particles with prescribed rules of collision and advection to replicate solutions of the Navier-Stokes equations. DPD and MPC accomplish this by introducing momentum carrying quasi-particles that fill the simulation volume and collide with immersed particles and each other, propagating momentum through a sequence of collisions, while LB tracks velocity and momentum distributions on a lattice and transfers momentum between adjacent points. These techniques can be applied to over-damped motion of micro-particles  characterized by low Reynolds numbers (Stokes flow).  However, the need to stably resolve the dynamics of the fluid on time scales short relative to momentum relaxation of the fluid mean that these calculations are more challenging to deploy for large numbers of micro-particles.\cite{pse}.
	 
	 The Fluctuating Immersed Boundary (FIB)\cite{fib} and Force Coupling Method (FCM)\cite{fcm} belong to a class of explicit solvent simulations that numerically solve the Stokes (or Navier-Stokes) equations in the computational domain using, for example, finite difference, volume, or Fourier decomposition techniques. Both methods approximate the particles as a collection of point forces distributed in the fluid, and much work has focused on developing fast and accurate kernels to construct these point distributions. Methods like FCM and FIB are most accurate for dilute suspensions where the mean particle separation is relatively large, and higher accuracy methods are not necessarily straightforward to derive. For example, the near-field or short-range hydrodynamic interactions between nearly touching particles are controlled in FCM/FIB by the choice of particle kernels, so lubrication forces cannot directly be accounted for. Furthermore, in the dilute suspension limit where they are most accurate, the FCM and FIB methods become less efficient and require increasingly large amounts of computer memory to perform the calculation. 
	
	Implicit solvent approaches to simulations with multi-body HI are formulated using the Green's function to represent the flows produced by particles and Fax\'{e}n laws to infer the motion induced by external flows impinging on their surfaces, respectively. The most common implicit solvent method is the Rotne-Prager-Yamakawa (RPY) mobility coupling\cite{rpy}, in which the first order interactions (force monopole and isotropic quadrupole) for spherical particles are summed pair-wise.  Three-body and higher-order multi-body effects are neglected to first-order. High order expansions follow directly from this level of approximation by systematically extending the multipole expansion. For example, the Stokesian Dynamics (SD) method extends RPY by adding the dipole and isotropic part of the octopole moment to the multipole expansion along with constraints on the hydrodynamic forces between two nearly touching particles\cite{sd}. High order multipole expansions contain embedded constraints, the solution to which accounts for the many-body interactions. The first such constraint appears with the force dipole, and its physical interpretation arises from the rigid body motion of immersed particles. Rigid particles cannot deform with the local fluid strain, and as a result, to first order, particles acquire a symmetric force dipole that produces additional disturbance flows on top of any imposed flow field to resist this deformation \cite{kimkarrila}. The symmetric force dipole is called the stresslet, so the rigid particle constraint is called the stresslet constraint in the context of hydrodynamic interactions. 
	
	The stresslet constraint increases viscous dissipation in the fluid and is therefore required to accurately compute many solution transport properties. For example, the RPY formulation predicts that the short-time self-diffusivity is independent of the volume fraction of suspended components.  However, when the stresslet constraint is added to the RPY formulation, the volume fraction dependence of the short-time self-diffusivity is recovered. Other types of dynamic constraints can arise from the computational representation of the particles, for example composite-bead particles\cite{gang}, which tesselate the surface of an arbitrary shape with rigidly constrained RPY beads which acts as a low order but convergent approximation of the exact solution to the Stokes equations. In this case a set of constraint forces, which guarantee that beads within the same composite particle move as a rigid body, must be determined at each point in the composite-bead particle trajectory. The composite-bead approach is not dissimilar to the boundary integral or immersed boundary representations of Stokes flow\cite{pozrikidis}, where particle surfaces are represented as collections of point forces and dipoles. Higher accuracy representations of the fluid flow in Green's function approaches introduce additional constraints on higher order force multipoles. Such expansions are difficult to truncate accurately for nearly touching spheres. Therefore, in the SD framework the relative motion of nearly touching particles is constrained by a force balance which includes a model of the near-field forces produced by lubrication flows\cite{sd} and also serves to increase local viscous dissipation.  

We have recently presented a novel technique for rapid Brownian dynamics (BD) simulations \cite{pse} with HI incorporated at the Rotne-Prager-Yamakawa (RPY) level of approximation\cite{rpy}. This algorithm is called the Positively-Split Ewald (PSE) algorithm.  The HI induce a linear coupling between particle velocities and forces. This coupling is represented by the mobility tensor $\mathcal{M}$, which is decomposed via Ewald summation\cite{hasimoto} into a sparse, local, real space contribution, $\mathcal{M}_{R}$ and low-rank, global, wave space contribution, $\mathcal{M}_{W}$.   The PSE formulation is unique because both $\mathcal{M}_{R}$ and $\mathcal{M}_{W}$ are independently symmetric and positive semi-definite for all particle configurations. This property is fundamental to the PSE algorithm and is used to sample the stochastic displacements associated with Brownian motion from the two independent normal distributions with covariances defined by $\mathcal{M}_{R}$ and $\mathcal{M}_{W}$. 
	
Clearly, additional hydrodynamic constraints must be added to RPY simulations to accurately capture the dynamics and transport properties of complex fluids.   Here, we describe an extension of the PSE algorithm to a representation that accounts for additional multipole moments corresponding to the hydrodynamic torque and stresslet. The hydrodynamic stresslet is the first induced multipole moment and determining its value requires solving a linear system of equations derived from the so-called stresslet constraint.  Approaches to handle this constraint with the PSE method are discussed in detail and are applicable to a broader class of constrained hydrodynamic equations. 
In Section \ref{sec:MobilityFormulation} we present a novel formulation of the mobility tensor including the particle motions induced by forces, torques, and stresslets, and discuss the positively-split Ewald sum of the new mobility tensor. Section \ref{sec:EwaldSum} details the accelerated Ewald sum procedure used to rapdly evaluate the mobility tensor. Section \ref{sec:Sampling} addresses the stresslet-constrained particle equations of motion and presents a method to sample Brownian displacements consistent with the constrained particle motions. Finally, we benchmark the performance of our method and assess its accuracy in several test cases. 
	
\section{ Formulation of the Positively-Split Mobility Tensor }
\label{sec:MobilityFormulation}

	\subsection{The Mobility Tensor}
	\label{sec:MobilityTensor}
	
The small length scale associated with colloidal particles means that local fluid motion is described by the Stokes equations. As particles move through the fluid they propagate flows which entrain other particles. This fluid-mediated coupling between immersed particles is represented by the mobility tensor $\mathcal{M}$, which is symmetric and positive semi-definite. Following the approach in the original PSE method \cite{pse}, $\mathcal{M}$ is constructed starting with the integral formulation of the Stokes equations for rigid suspended bodies\cite{ladyzhenskaya} 
	\begin{align}
		u_{i}\left({\bf x}\right) &= \int_{S_{\beta}} dS_{\beta}\left({\bf y}\right) \, J_{im}\left({\bf x},{\bf y}\right) f_{m}\left({\bf y}\right), \label{eqn:mpole1}
	\end{align}
	where $u_{i}\left({\bf x}\right)$ is the $i^{th}$ component of the fluid velocity at a point ${\bf x}$ in the fluid propagated by the force density ${\bf f}\left({\bf y}\right)$ on the particle surface $S_{\beta}$ through the Green's function ${\bf J}$. In this and following equations, Einstein summation notation is use to represent operations on vector and tensor quantities. The particle are assumed to be spherical and their force density is represented by a local expansion about the particle center using the multipole moments
	\begin{align}
		f_{j} &= \frac{1}{4\pi a^{2}} \, F_{j}   + \frac{3}{4\pi a^{3}} \, \hat{n}_{n} \, C_{nj} + \cdots, \label{eqn:mpole2}
	\end{align}				
	where ${\bf F}$ is the total force on the particle and ${\bf C}$ is the traceless couplet, whose symmetric and antisymmetric components are the stresslet and torque, respectively, and the expansion is truncated after the couplet. The vector $\hat{\bf n}$ is the normal to the particle surface and $a$ is the particle radius. The force and couplet are defined by
	\begin{align}
		F_{m} &= \int_{S_{\beta}}dS_{\beta} \, f_{m} , \label{eqn:mpole3}\\
		C_{mn} &= \int_{S_{\beta}}dS_{\beta} \, \Big[ \left(y_{m}-x_{\beta,m}\right)f_{n} -  \label{eqn:mpole4}\\
			   &\qquad \frac{1}{3} \delta_{mn}\left(y_{p}-x_{\beta,p}\right)f_{p} \Big] . \nonumber
	\end{align}				
	Particulate motion is inferred from the flow field impinging on a particle using Fax\'{e}n laws, which dictate how a particle moves in response to the stresses that a flow field exerts on its surface
	\begin{align}
		U_{i} &= \frac{1}{4\pi a^{2}} \int_{S_{\alpha}} dS_{\alpha}\left({\bf x}\right) \, u_{i}\left({\bf x}\right), \label{eqn:fax1}\\
		D_{ij} &= \frac{3}{4\pi a^{3}} \int_{V_{\alpha}} dV_{\alpha}\left({\bf x}\right) \, \nabla_{j} \, u_{i},\label{eqn:fax2a}
	\end{align}
	where $S_{\alpha}$ is the surface of particle $\alpha$, and $V_{\alpha}$ is its volume. The particle velocity gradient, ${\bf D}$, is composed of a symmetric part, the rate of strain ${\bf E}$, and an anti-symmetric part, the angular velocity ${\boldsymbol \Omega}$. The volume integral in \eqref{eqn:fax2a} can be rewritten using the Divergence Theorem
	\begin{align}
		D_{ij} &= \frac{3}{4\pi a^{3}} \int_{S_{\alpha}} dS_{\alpha}\left({\bf x}\right) \, u_{i} \, \hat{n}_{j} ,\label{eqn:fax2b}
	\end{align}
	so that all equations are written in terms of surface integrals.
			
	The linearity of the Stokes equations allows independent coupling and superposition of the total force or couplet to the velocity or velocity gradient. Substituting the fluid velocity \eqref{eqn:mpole1} and the multipole expansion \eqref{eqn:mpole2} into the Fax\'{e}n laws \eqref{eqn:fax1},\eqref{eqn:fax2b} gives the following relations between the force moments and the velocity and its gradient
	\begin{align}
		U_{i}^{\alpha} &= M_{UF,im}^{\alpha\beta} \, F_{m}^{\beta} + M_{UC,imn}^{\alpha\beta} \, C_{mn}^{\beta}, \\
		D_{ij}^{\alpha} &= M_{DF,ijm}^{\alpha\beta} \, F_{m}^{\beta} + M_{DC,ijmn}^{\alpha\beta} \, C_{mn}^{\beta},
	\end{align}				
	where the coupling pair mobility tensors are given in terms of integrals of the Green's function over the particle surfaces
	\begin{widetext}
	\begin{align}
		M_{UF,im}^{\alpha\beta} &= \frac{1}{4 \pi a^{2}} \, \int_{S_{\alpha}}dS_{\alpha}\left({\bf x}\right) \, \frac{1}{4\pi a^{2}} \, \int_{S_{\beta}} dS_{\beta}\left({\bf y}\right) \, J_{im}\left({\bf x},{\bf y}\right),  \label{eqn:UF}\\
		M_{UC,imn}^{\alpha\beta} &= \frac{1}{4 \pi a^{2}} \, \int_{S_{\alpha}}dS_{\alpha}\left({\bf x}\right) \, \frac{3}{4\pi a^{3}} \, \int_{S_{\beta}} dS_{\beta}\left({\bf y}\right) \, J_{im}\left({\bf x},{\bf y}\right) \, \hat{n}^{\beta}_{n}, \label{eqn:UC} \\
		M_{DF,ijm}^{\alpha\beta} &= \frac{3}{4 \pi a^{3}} \, \int_{S_{\alpha}} dS_{\alpha}\left({\bf x}\right) \, \hat{n}^{\alpha}_{j} \, \frac{1}{4\pi a^{2}} \, \int_{S_{\beta}} dS_{\beta}\left({\bf y}\right) \, J_{im}\left({\bf x},{\bf y}\right), \label{eqn:DF} \\
		M_{DC,ijmn}^{\alpha\beta} &= \frac{3}{4 \pi a^{3}} \, \int_{S_{\alpha}} dS_{\alpha}\left({\bf x}\right) \, \hat{n}^{\alpha}_{j} \, \frac{3}{4\pi a^{3}} \, \int_{S_{\beta}} dS_{\beta}\left({\bf y}\right) \, J_{im}\left({\bf x},{\bf y}\right) \, \hat{n}^{\beta}_{n} . \label{eqn:DC}
	\end{align}
	\end{widetext}
	For periodic systems, which are commonly used in simulation, the appropriate Green's function is given by \citeauthor{hasimoto}\cite{hasimoto},
	\begin{equation}
		J_{im}\left({\bf x},{\bf y}\right) = \frac{1}{\eta V}\sum\limits_{{\bf k}\neq{\bf 0}} e^{i{\bf k}\cdot\left({\bf x}-{\bf y}\right)} \frac{1}{k^{2}}  \left( \delta_{im} - \hat{k}_{i}\hat{k}_{m} \right), \label{eqn:hasimoto}
	\end{equation}
	where ${\bf J}\left({\bf x},{\bf y}\right) \cdot {\bf F}$ is the fluid velocity at point ${\bf x}$ propagated by a force ${\bf F}$ located at point ${\bf y}$ and its periodic images on a lattice, $\delta_{im}$ is the identity tensor, ${\bf k}$ are drawn from the set of reciprocal space lattice vectors excluding the zero vector, $\eta$ is fluid viscosity, and $V$ is the periodic cell volume. Substituting equation \eqref{eqn:hasimoto} into equations \eqref{eqn:UF}-\eqref{eqn:DC} and evaluating the surface integrals in Fourier space gives algebraic expressions for the pair mobilities of spherical particles with equal, finite size
	\begin{widetext}
	\begin{align}
		M_{UF,im}^{\alpha\beta} &= \frac{1}{\eta V}\sum\limits_{{\bf k}\neq{\bf 0}} e^{i{\bf k}\cdot\left({\bf x}_{\alpha}-{\bf x}_{\beta}\right)} \frac{1}{k^{2}} \left( \frac{\sin k a}{k a} \right)^{2}  \left( \delta_{im} - \hat{k}_{i}\hat{k}_{m} \right), \label{eqn:MUF} \\
		M_{UC,imn}^{\alpha\beta} &= -\frac{1}{\eta V}\sum\limits_{{\bf k}\neq{\bf 0}} e^{i{\bf k}\cdot\left({\bf x}_{\alpha}-{\bf x}_{\beta}\right)} \frac{1}{k^{2}} \left( \frac{\sin k a}{k a} \right) \, \left( 3\sqrt{-1} \, \frac{\sin k a - k a \cos k a}{k^{3}a^{3}} \right) \, \left( \delta_{im}k_{n} - \hat{k}_{i}\hat{k}_{m}k_{n} \right), \label{eqn:MUC} \\
		M_{DF,ijm}^{\alpha\beta} &= \frac{1}{\eta V}\sum\limits_{{\bf k}\neq{\bf 0}} e^{i{\bf k}\cdot\left({\bf x}_{\alpha}-{\bf x}_{\beta}\right)} \frac{1}{k^{2}} \left( 3\sqrt{-1} \,  \frac{\sin k a - k a \cos k a}{k^{3}a^{3}} \right) \, \left( \frac{\sin k a}{k a} \right) \, \left( \delta_{im}k_{j} - \hat{k}_{i}\hat{k}_{m}k_{j} \right), \label{eqn:MDF} \\
		M_{DC,ijmn}^{\alpha\beta} &= -\frac{1}{\eta V}\sum\limits_{{\bf k}\neq{\bf 0}} e^{i{\bf k}\cdot\left({\bf x}_{\alpha}-{\bf x}_{\beta}\right)} \frac{1}{k^{2}} \left( 3\sqrt{-1} \, \frac{\sin k a - k a \cos k a}{k^{3}a^{3}} \right)^{2} \, \left( \delta_{im}k_{j}k_{n} - \hat{k}_{i}\hat{k}_{m}k_{j}k_{n} \right). \label{eqn:MDC}
	\end{align}	
	\end{widetext}
	The relations given in Equations \eqref{eqn:MUF}-\eqref{eqn:MDC} produce a pair mobility tensor,
	\begin{align}
		{\bf M}^{\alpha\beta} = \begin{bmatrix} {\bf M}_{\rm UF}^{\alpha\beta} & {\bf M}_{\rm UC}^{\alpha\beta} \\ {\bf M}_{\rm DF}^{\alpha\beta} & {\bf M}_{\rm DC}^{\alpha\beta} \end{bmatrix}, \label{eqn:pairmob}
	\end{align}	
	that is symmetric and positive semi-definite for all particle separations with no special regularization required. This is a stark constrast to past approaches, for example the original work by \citeauthor{rpy}\cite{rpy} and more recent efforts by \citeauthor{RPY_Periodic_Shear}\cite{RPY_Periodic_Shear}, which require piecewise construction of the pair mobility tensors with special regularization assertions for overlapping particles. 
	In the approach presented here, the coupling between arbitrary force moments and arbitrary velocity moments is expressed simply in Fourier space as the point force solution multiplied by the product of size-dependent shape factors. The coupling between the $p^{th}$ velocity moment and the $q^{th}$ force moment (where $p=0$ corresponds to the linear velocity and $q=0$ corresponds to the force) is \cite{ladd_mobility}
	\begin{widetext}
	\begin{equation}
		M_{pq}^{\alpha\beta} = \left(i\right)^{p-q} \left( 2p + 1 \right)!! \, \left( 2q + 1 \right)!! \, \sum\limits_{{\bf k}\neq{\bf 0}} \, e^{i{\bf k}\cdot\left({\bf x}_{\alpha}-{\bf x}_{\beta}\right)} \, \frac{j_{p}\left(ka\right)\,j_{q}\left(ka\right)}{\eta V k^{2}} \, \overline{\hat{{\bf k}}^{p}} \, \left( {\bf I} - \hat{\bf k}\hat{\bf k} \right) \, \overline{\hat{{\bf k}}^{q}},
	\end{equation}
	\end{widetext}
	where $j_{p}$ is the spherical Bessel function of the first kind of order $p$, and $\overline{\hat{{\bf k}}^{p}}$ denotes an irreducible tensor of rank $p$. 
	The shape factors are tensorial functions that describe the way in which objects of a given shape, in this case spheres, propagate flows into the fluid. 
	The shape factor associated with $U$ and $F$ is $\sin\left(ka\right)/ka$, and the shape factor associated with $D$ and $C$ is $3\sqrt{-1}\,\left( \sin k a - k a \cos k a \right)/k^{3}a^{3} \, {\bf k}$. 
	
	\subsection{Positive Splitting of Mobility Tensor}
	\label{sec:MobilitySplitting}
	Here, we extend the previously reported PSE algorithm \cite{pse} to the mobility tensor including the couplet and velocity gradient. A representation of these tensors whose real space and wave space contributions to the Ewald sum are independently symmetric and positive definite, i.e. a representation that can be positively split, is necessary to rapidly sample the Brownian displacements using the split sampling of PSE technique. We begin by applying \citeauthor{hasimoto}'s sum splitting\cite{hasimoto}, which introduces the splitting function, $H(k,\xi) = \left( 1 + k^{2}/4\xi^{2}\right) \, e^{-k^{2}/4\xi^{2}}$, to decompose equations \eqref{eqn:MUF}-\eqref{eqn:MDC} into a long-ranged wave-space part ${\bf M}^{(w)}$ and a short-ranged real-space part ${\bf M}^{(r)}$ as ${\bf M} = {\bf M}^{(r)} + {\bf M}^{(w)}$ yields
	\begin{widetext}
	\begin{align}
		M_{UF,im}^{\alpha\beta,(w)} &= \frac{1}{\eta V}\sum\limits_{{\bf k}\neq{\bf 0}} e^{i{\bf k}\cdot\left({\bf x}_{\alpha}-{\bf x}_{\beta}\right)} \frac{1}{k^{2}} \left( \frac{\sin k a}{k a} \right)^{2}  \, H\left(k,\xi\right) \, \left( \delta_{im} - \hat{k}_{i}\hat{k}_{m} \right), \label{eqn:MUFw} \\
		M_{UC,imn}^{\alpha\beta,(w)} &= -\frac{1}{\eta V}\sum\limits_{{\bf k}\neq{\bf 0}} e^{i{\bf k}\cdot\left({\bf x}_{\alpha}-{\bf x}_{\beta}\right)} \frac{1}{k^{2}} \left( \frac{\sin k a}{k a} \right) \, \left( 3i \frac{\sin k a - k a \cos k a}{k^{3}a^{3}} \right) \, H\left(k,\xi\right) \,  \left( \delta_{im}k_{n} - \hat{k}_{i}\hat{k}_{m}k_{n} \right), \label{eqn:MUCw} \\
		M_{DF,ijm}^{\alpha\beta,(w)} &= \frac{1}{\eta V}\sum\limits_{{\bf k}\neq{\bf 0}} e^{i{\bf k}\cdot\left({\bf x}_{\alpha}-{\bf x}_{\beta}\right)} \frac{1}{k^{2}} \left( 3i \frac{\sin k a - k a \cos k a}{k^{3}a^{3}} \right) \, \left( \frac{\sin k a}{k a} \right) \, H\left(k,\xi\right) \,  \left( \delta_{im}k_{j} - \hat{k}_{i}\hat{k}_{m}k_{j} \right), \label{eqn:MDFw} \\
		M_{DC,ijmn}^{\alpha\beta,(w)} &= -\frac{1}{\eta V}\sum\limits_{{\bf k}\neq{\bf 0}} e^{i{\bf k}\cdot\left({\bf x}_{\alpha}-{\bf x}_{\beta}\right)} \frac{1}{k^{2}} \left( 3i \frac{\sin k a - k a \cos k a}{k^{3}a^{3}} \right)^{2} \, H\left(k,\xi\right) \,  \left( \delta_{im}k_{j}k_{n} - \hat{k}_{i}\hat{k}_{m}k_{j}k_{n} \right), \label{eqn:MDCw}
	\end{align}	
	\end{widetext}
	and
	\begin{align}
		M_{UF,im}^{\alpha\beta,(r)}  &= F_{1} \, \left( \delta_{im} - \hat{r}_{i}\hat{r}_{m} \right) + F_{2} \, \hat{r}_{i}\hat{r}_{m}, \label{eqn:MUFr}
	\end{align}
	\begin{align}
		M_{UC,imn}^{\alpha\beta,(r)} &= G_{1} \, \left( \delta_{im}\hat{r}_{n} - \hat{r}_{i}\hat{r}_{m}\hat{r}_{m} \right) \label{eqn:MUCr} \\
									&\qquad + G_{2} \, \left( \delta_{in}\hat{r}_{m} + \delta_{mn}\hat{r}_{i} - 4\hat{r}_{i}\hat{r}_{m}\hat{r}_{n} \right), \nonumber 
	\end{align}
	\begin{align}
		M_{DC,ijmn}^{\alpha\beta,(r)} &= K_{1} \left( \delta_{ij}\delta_{mn} + \delta_{im}\delta_{jn} -4 \delta_{in}\delta_{jm} \right) \label{eqn:MDCr} \\
										&\qquad + K_{2}\left( \delta_{jm}\hat{r}_{i}\hat{r}_{n} - \hat{r}_{i}\hat{r}_{j}\hat{r}_{m}\hat{r}_{n} \right) \nonumber \\
				  						&\qquad + K_{3} \left( \delta_{ij}\hat{r}_{m}\hat{r}_{n} + \delta_{im}\hat{r}_{j}\hat{r}_{n} + \delta_{jn}\hat{r}_{i}\hat{r}_{m} \right. \nonumber \\
				  						&\qquad \left. + \delta_{mn}\hat{r}_{i}\hat{r}_{j} + \delta_{in}\hat{r}_{j}\hat{r}_{m} - 6\hat{r}_{i}\hat{r}_{j}\hat{r}_{m}\hat{r}_{n} \right. \nonumber \\
				  						&\qquad \left. - \delta_{in}\delta_{jm} \right), \nonumber
	\end{align}
	where
	\begin{equation}
		H\left(k,\xi\right) = \left( 1 + \frac{k^{2}}{4\xi^{2}} \right) e^{-k^{2}/4\xi^{2}},
	\end{equation}
	and $F_{1}$, $F_{2}$, $G_{1}$, $G_{2}$, $K_{1}$, $K_{2}$, and $K_{3}$ are exponentially decaying scalar functions of $r/a$ and $\xi\,a$ given in Appendix \ref{app:AppendixA}. The real space mobility functions are calculated by analytically evaluating the inverse Fourier transform of ${\bf M} - {\bf M}^{(w)}$, and representing the tensors in a minimal form using the symmetry and tracelessness of the various mobility couplings, where appropriate. 
	
	The splitting parameter, $\xi$, defines a length scale ($\xi^{-1}$) that determines the partitioning of the total sum between the real space and wave space contributions. The Ewald sum technique filters high-frequency modes of motion with wavelengths shorter than $\xi^{-1}$ out of the wave space sum and reconstructs them explicitly in real space. Certain filters, such as \citeauthor{hasimoto}'s splitting function, guarantee that the magnitude of the terms in both the real space and wave space sums decay exponentially with distance and wave-vector, respectively \cite{hasimoto}. Because the terms in the sum decay so rapidly, $\xi$ controls the locality of the interactions included in the real space sum; a given particle pair will only have an appreciable contribution to the sum if the distance between the particles is less than $\xi^{-1}$ because at larger distances the propagated flows have decayed to negligible magnitudes. This length scale $\xi^{-1}$ can be freely chosen to minimize the total computational effort by matching it to physically relevant length scales such as the blob size of a polymer suspension or the correlation length of a colloidal gel network. 
	
	The pair mobility tensor in equation \eqref{eqn:pairmob} constructed from the couplings defined in equations \eqref{eqn:MUF}-\eqref{eqn:MDC}, combined with \citeauthor{hasimoto}'s sum splitting, guarantee that the real space and wave space contributions, ${\bf M}^{\alpha\beta} = {\bf M}^{\alpha\beta,(w)} + {\bf M}^{\alpha\beta,(r)}$, are independently symmetric and positive semi-definite regardless of particle configuration and splitting parameter. A proof of the positive-definiteness of the Ewald sum is given in Appendix \ref{app:AppendixC}.
	
	The pair mobilities can easily be extended to many-body simulations by summing the pairwise interactions for each particle
	\begin{align}
		\begin{bmatrix}
			{\bf U}^{\alpha} \\ {\bf D}^{\beta}
		\end{bmatrix} &= \sum\limits_{\beta} \left( \begin{bmatrix}
			{\bf M}_{\rm UF} & {\bf M}_{\rm UC} \\ {\bf M}_{\rm DF} & {\bf M}_{\rm DC}
\end{bmatrix}^{\alpha\beta,(w)}	 + \right. \label{eqn:pairsums} \\
					&\qquad \left. \begin{bmatrix}
			{\bf M}_{\rm UF} & {\bf M}_{\rm UC} \\ {\bf M}_{\rm DF} & {\bf M}_{\rm DC}
\end{bmatrix}^{\alpha\beta,(r)} \right) \cdot \begin{bmatrix}
			{\bf F}^{\beta} \\ {\bf C}^{\beta}
		\end{bmatrix} \nonumber. 
	\end{align}
	This set of sums can be written using the grand mobility tensor $\mathcal{M}$ or its Ewald sum $\mathcal{M} = \mathcal{M}_{w} + \mathcal{M}_{r}$,
	\begin{equation}
		\begin{bmatrix}
			{\bf U} \\ {\bf D}
		\end{bmatrix} = \mathcal{M} \cdot \begin{bmatrix}
			{\bf F} \\ {\bf C}
		\end{bmatrix} = \left( \mathcal{M}_{w} + \mathcal{M}_{r} \right) \cdot \begin{bmatrix}
			{\bf F} \\ {\bf C}
		\end{bmatrix},
	\end{equation}
	where ${\bf U}$ and ${\bf D}$ are vectors containing all particle velocities and velocity gradients, respectively, and ${\bf F}$ and ${\bf C}$ respectively contain all particle forces and couplets. The grand mobility tensor, $\mathcal{M}$, constructed using the pairwise mobility tensors presented above is symmetric and positive semi-definite for all particles configurations, as are $\mathcal{M}_{w}$ and $\mathcal{M}_{r}$. 
		
		\subsubsection{Mobility Relation for Rigid Particles}	
		\label{sec:RigidMobility}	
		
		A more common representation of the mobility problem, and one that will be useful in the context of rigid particles, decomposes both the velocity gradient and the couplet into their symmetric and antisymmetric parts
		\begin{equation}
			\begin{bmatrix}
				\mathcal{U} \\ {\bf E}
			\end{bmatrix} = \mathcal{M} \cdot \begin{bmatrix}
				\mathcal{F} \\ {\bf S}
			\end{bmatrix},
		\end{equation}
		where $\mathcal{U}$ is a column vector containing both the linear and angular velocity of all particles, $\mathcal{U} = [ {\bf U}^{T} \:\: \boldsymbol{\Omega}^{T} ]^{T}$, $\mathcal{F}$ is a column vector containing the force and torque on all particles, $\mathcal{F} = [ {\bf F}^{T} \:\: {\bf L}^{T} ]^{T}$ and ${\bf E}$ and ${\bf S}$ are the rate of strain and stresslet for all particles, respectively. These quantities are defined for the $\alpha^{th}$ particle according to
		\begin{align}
			\boldsymbol{\Omega}^{\alpha} &= \frac{1}{2} \, \boldsymbol{\epsilon} : {\bf D}^{\alpha}, \\
			{\bf E}^{\alpha} &= \frac{1}{2} \, \left( {\bf D}^{\alpha} + \left({\bf D}^{\alpha}\right)^{T} \right), \\ 
			{\bf L}^{\alpha} &= \frac{1}{2} \, \boldsymbol{\epsilon} : \left( {\bf C}^{\alpha} - \left({\bf C}^{\alpha}\right)^{T} \right), \\
			{\bf S}^{\alpha} &= \frac{1}{2} \, \left( {\bf C}^{\alpha} + \left({\bf C}^{\alpha}\right)^{T} \right),
		\end{align}	
		where $\boldsymbol{\epsilon}$ is the Levi-Civita tensor. When subject to straining flows, rigid particles acquire a stresslet whose value is such that the local rate of strain is zero,
		\begin{equation}
			\begin{bmatrix}
				\mathcal{U} \\ {\bf 0}
			\end{bmatrix} =  \begin{bmatrix}
				{\bf M}_{\mathcal{UF}} & {\bf M}_{\mathcal{U}{\rm S}} \\ {\bf M}_{{\rm E}\mathcal{F}} & {\bf M}_{\rm ES}
\end{bmatrix} \cdot \begin{bmatrix}
				\mathcal{F} \\ {\bf S}
			\end{bmatrix}. \label{eqn:rigidconstraint}
		\end{equation}
		Enforcing this condition in simulations requires solving the linear system of equations \eqref{eqn:rigidconstraint} for the unknown particle velocities $\mathcal{U}$ and stresslets ${\bf S}$. The solution to \eqref{eqn:rigidconstraint} is
		\begin{align}
			\mathcal{U} &= \left( \mathcal{M}_{\mathcal{UF}} - \mathcal{M}_{\mathcal{U}{\rm S}} \cdot \mathcal{M}_{\rm ES}^{-1} \cdot \mathcal{M}_{{\rm E}{\mathcal F}} \right) \cdot \mathcal{F} \label{eqn:constrained_vel}\\
			{\bf S} &= -\mathcal{M}_{\rm ES}^{-1} \cdot \mathcal{M}_{{\rm E}\mathcal{F}} \cdot \mathcal{F},
		\end{align}
		where the quantity $\mathcal{M}_{\mathcal{UF}} - \mathcal{M}_{\mathcal{U}{\rm S}} \cdot \mathcal{M}_{\rm ES}^{-1} \cdot \mathcal{M}_{{\rm E}{\mathcal F}}$ is the stresslet-constrained mobility tensor, which describes the coupling between particle forces and particle velocities as modified by the stresslet constraint. 
		
	\subsection{ Efficient Evaluation of the Ewald Sum }
	\label{sec:EwaldSum}
	For a given representation of the grand Mobility tensor, different techniques can be used to evaluate the real space and wave space parts of the Ewald sum. Careful selection of these techniques to leverage the unique structure of each sum can accelerate the calculations and reduce the computational complexity while maintaining calculation accuracy. Because the terms in both the real space and wave space sum both decay rapidly, the sums can be truncated at a fixed number of terms. Here we describe two fast algorithms to evaluate the truncated sums, one for the real space sum and one for the wave space sum, that have a linear scaling of the computational cost in the number of particles modeled.
	
		\subsubsection{ Real Space Calculations }
		\label{sec:EwaldSumReal}
		The real space contribution to the Ewald summation contains only terms that decay exponentially with the distance between the particles. The rate of the decay is controlled by the splitting parameter, $\xi$, so that the strength of the interaction between two particles separated by a distance $r$ is approximately $\epsilon_{r} \sim e^{-\xi^{2} \, r^{2} }$. Therefore, only particles whose distance is on the order of $\xi^{-1}$ or smaller propagate flows that are strong enough to effect one another. One can define a cutoff radius, $r_{\rm cut}$, beyond which the interactions are ignored, and incur a finite error of magnitude $\epsilon_{r} \sim e^{-\xi^{2} \, r_{\rm cut}^{2} }$. Summing local interactions in this way is readily handled using linked-cell lists\cite{allen_tildesley} to divide the simulation cell into sub-cells of size $r_{\rm cut} \sim \xi^{-1}$ and only considering interactions between particles in the same or adjacent cells. In this scheme, on average each particle participates in ${\phi}\,r_{\rm cut}^{d_{f}}$ pair interactions, where $\phi$ is the particle volume fraction and $d_{f}$ is the fractal dimension of the particle dispersion. The number of interactions each particle has is independent of the total number of particles in the system, so the overall computational cost of summing the interactions for all $N$ particles is $O(N \, {\phi}\,r_{\rm cut}^{d_{f}})$, which scales linearly with the total number of particles. 
		
		\subsubsection{ Wave Space Calculations }
		\label{sec:EwaldSumWave}
		The wave space component of the mobility can be written as a sequence of linear operations
		\begin{equation}
			\mathcal{M}_{w} = \mathbb{D}^{\dagger} \cdot {\bf P}^{\dagger} \cdot {\bf B} \cdot {\bf P} \cdot \mathbb{D}. \label{eqn:WSops}
		\end{equation}
		$\mathbb{D}$ is the non-uniform Discrete Fourier transform (NUDFT) operator. The application of $\mathbb{D}$ transforms the point moments applied on the particles into Fourier space. The inverse NUDFT, $\mathbb{D}^{\dagger}$ converts the wave space representation of the velocity moments to real space by evaluating the velocities at the particle positions. ${\bf B}$ is a symmetric block diagonal matrix whose non-zero entries are
		\begin{equation}
			{\bf B}_{ii} = \frac{ H(k_{i},\xi) }{ {\eta}Vk_{i}^{2} } \, \left( {\bf I}-\hat{\bf k}_{i}\hat{\bf k}_{i} \right).
		\end{equation}
		The application of ${\bf B}$ multiplies the Fourier components by the Hasimoto factor $H(k,\xi)/{\eta}Vk^{2}$ and projects the wave space representation of the force onto the space of divergence-free velocity fields via the projector $({\bf I}-\hat{\bf k}\hat{\bf k})$. The operator ${\bf P}$ is a wave space projector containing non-negative factors that maps the Fourier space force moments onto the Fourier space force density, and ${\bf P}^{\dagger}$ maps the Fourier space velocity field to the Fourier space velocity moments. ${\bf P}$ is a non-square matrix with non-zero rectangular blocks along the main diagonal
		\begin{equation}
			{\bf P}_{ii} = \begin{bmatrix} \frac{\sin{k_{i}a}}{k_{i}a} & 3 \, \sqrt{-1} \, \frac{\sin{k_{i}a}-k_{i}a\,\cos{k_{i}a}}{k_{i}^{3}a^{3}}\,{\bf k}_{i} \end{bmatrix}.
		\end{equation}
		The product ${\bf P}_{ii}^{\dagger} \cdot {\bf B}_{ii} \cdot {\bf P}_{ii}$ gives the wave space representation of the mobility tensor at the $i^{\rm th}$ wave vector, i.e. the summands in Equations \eqref{eqn:MUFw}-\eqref{eqn:MDCw} without the exponential factors. The exponential factors and the sum over ${\bf k}\neq{\bf 0}$ come from the forward and inverse NUDFT, $\mathbb{D}$ and $\mathbb{D}^{\dagger}$. 
		
		To accelerate the evaluation of the wave space sum \eqref{eqn:WSops}, we follow the approach taken for Stokeslets in the spectral Ewald (SE) method of \citeauthor{lindbo-tornberg}\cite{lindbo-tornberg}, using the non-uniform FFT (NUFFT) to obtain the action of $\mathbb{D}$ and $\mathbb{D}^{\dagger}$. The SE approach is described in detail in Ref. \cite{lindbo-tornberg}, but the salient points are summarized here. 
		The fundamental idea behind the NUFFT is to evaluate the wave space sums \eqref{eqn:MUFw}-\eqref{eqn:MDCw} on a regular grid so as to be able to leverage the FFT algorithm. Direct evaluation of the sums in \eqref{eqn:MUFw}-\eqref{eqn:MDCw} over $N_{k}$ wave-vectors requires $\mathcal{O}\left(N_{k}^{2}\right)$ operations, while the FFT algorithm requires only $\mathcal{O}\left(N_{k}\,\log N_{k}\right)$ operations.
		
		 Particle positions in dynamic simulations do not in general coincide with points on a uniform grid, so information must be transferred between the particle positions and nearby points on the regular grid in order to apply the FFT algorithm. In particular, particle force moments need to be spread onto the grid in order to determine the fluid response, and fluid velocities need to be interpolated from the grid to infer the resulting particle motions. The NUFFT method accomplishes this task using a Gaussian kernel function truncated at $m$ standard deviations to accomplish the spreading and interpolation. The value of $m$ is chosen to ensure that the Gaussian kernel is sufficiently resolved on the grid to control the error associated with the spreading and interpolation (hereafter referred to collectively as quadrature) operations to within a desired error tolerance. The quadrature operations are performed by evaluating convolutions of data with the Gaussian kernels on a subgrid of $P^{3}$ grid points centered on each particle, where $P$ is a sufficiently large integer for a given tolerance \cite{lindbo-tornberg}. The parameters $m$ and $P$ collectively control the decay rate of the Gaussian kernel. 
		
		Use of the Gaussian kernels corresponds to representing the NUDFT as $\mathbb{D} = \mathbb{F}\cdot{\bf Q}$, where $\mathbb{F}$ represents the FFT. The operator ${\bf Q}$ is a Gaussian kernel that spreads particle force moments to the uniform grid, and its adjoint ${\bf Q}^{\dagger} = \sigma\,{\bf Q}^{T}$ is the interpolation operator, where $\sigma$ is the volume of a grid cell. Therefore,
		\begin{equation}
			\mathcal{M}_{w} = \sigma\,{\bf Q}^{T}\cdot\mathbb{F}^{\dagger}\cdot{\bf P}^{\dagger}\cdot{\bf B}\cdot{\bf P}\cdot\mathbb{F}\cdot{\bf Q}, \label{eqn:WSops2}
		\end{equation}
		
		The operator ${\bf Q}$ defines the quadrature operations for point particles. In \eqref{eqn:WSops2} each component of the force and couplet is mapped to a separate FFT grid as though the particles were points through the application of ${\bf Q}$. The FFT of each of these components is evaluated independently by $\mathbb{F}$, and the wave space representation of all the force moments are projected down onto the (3-dimensional) wave space force density by the operator ${\bf P}$, which accounts for the finite size and spherical shape of each particle. The operator ${\bf B}$ maps the wave space force density into the wave space fluid velocity, which is then mapped to the real space linear velocity, angular velocity, and rate of strain by $\mathbb{F}^{\dagger}\cdot{\bf P}^{\dagger}$. Particle motion is interpolated from the resulting real space velocity components by $\sigma\,{\bf Q}^{T}$. This approach requires 11 FFT grids in total (3 for the independent components of the force and 8 for the independent components of the traceless couplet) and only works if all particles have the same shape factors, i.e. same size and shape. For simulations of monodisperse particle suspensions, we have found this approach to have the best performance due to the high efficiency of the FFT algorithm and the simplicity of the point particle quadrature operations. 
		
		For polydisperse particle suspensions, each particle can in principle have a different shape factor, so equation \eqref{eqn:WSops2} cannot be directly applied. It is possible to instead directly transfer each of the three independent components of the particle force density to a FFT grid, accounting for particle size effects in the process. Modified quadrature kernels that map the force and couplet onto the real space force density and project the force density onto the real space force density grid can be derived by applying the inverse Fourier transform to the shape factors in ${\bf P}$,
		\begin{equation}
			\mathcal{M}_{w} = \sigma \, \mathbb{Q}^{T} \cdot \mathbb{F}^{\dagger} \cdot {\bf B} \cdot \mathbb{F} \cdot \mathbb{Q}, \label{eqn:WSops3}
		\end{equation}
		where the operator $\mathbb{Q}$ represents the modified quadrature kernels. There is a different modified kernel for each force moment that projects it onto the force density, i.e. ${\bf Q}$ can be represented by a non-square matrix ${\bf Q} = \left[ {\bf Q}_{\mathcal{F}} \:\: {\bf Q}_{\rm S} \right]$, such that $\left[ {\bf Q}_{\mathcal{F}} \:\: {\bf Q}_{\rm S} \right]\cdot\left[ \mathcal{F} \:\: {\bf S} \right]^{T}$ is the projection of the generalized force and stresslet onto the force density representation on a uniform grid. Detailed discussion of such modified kernels is left for future work. 
		
		\subsubsection{ Wave Space Calculations on Deformed FFT Grids }
		\label{sec:DeformedGrid}	
		NUFFT methods, such as the one presented here, often use orthorhombic lattices to evalute the FFT. However, in dynamic simulations of flowing materials with periodic boundary conditions, the periodic lattice must deform. Two common examples of this are the Lees-Edwards \cite{lees-edwards} and Kraynik-Reinelt \cite{kraynik-reinelt} boundary conditions used to model shear and extensional flow, respectively. As the stresslet represents the first order correction to particle dynamics due to the rigidity constraint, it is natural to investigate particles in shear or extensional flows at the stresslet level of hydrodynamic approximation. 
		\begin{figure}
			\includegraphics{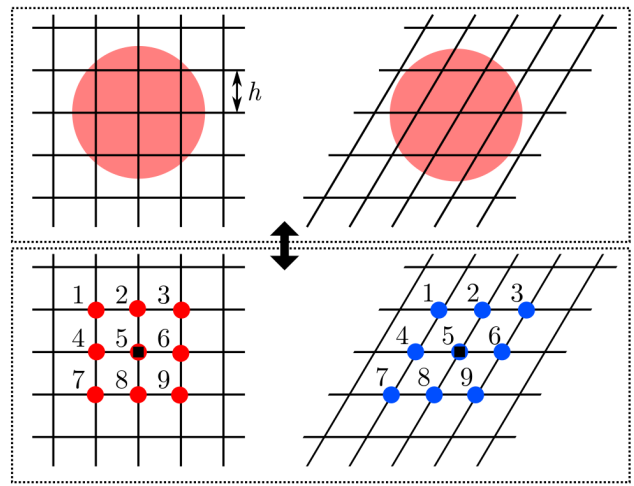}
			\caption{Illustrative schematic of the NUFFT quadrature operations in two dimensions, where the objective is to transfer information between the particle positions and the local lattice positions. (top) Particles positions relative to the underlying FFT lattice with spacing $h$. (bottom) Lattice points within the neighborhood of the particle, denoted with red circles on the orthorhombic lattice and with blue circles on the deformed lattice, and the location of the particle center, denoted with a black square. Lattice points are labeled with their index number. The points of the deformed lattice (right) have different positions relative to the particle center than the equivalent points in the orthorhombic lattice (left). In this example, the size of the particle's support in each dimesion, $P$, is 3. }
			\label{fig:quadrature}
		\end{figure}
		
		The difference between NUFFT methods applied to orthorhombic and non-orthorhombic (deformed) lattices is in the quadrature operations used to transfer information between the non-uniform particles positions and the uniform lattice points. The kernels used to transfer particle data to the grid depend on the relative positions of the grid points and the particle. Figure \ref{fig:quadrature} illustrates these differences for an orthorhombic lattice and the same lattice after undergoing linear shear. Points with the same index on different lattices carry different values of the Gaussian quadrature kernel because they are distributed around the particle center differently. To our knowledge, the effect of these differences have not been incorporated into the error bounds for NUFFT quadrature operations on deformed grids. Accurate error bounds that account for lattice deformation are essential to performing simulations with well-controlled numerical error. Here, we report quadrature error bounds for linearly deformed grids derived following the approach that \citeauthor{lindbo-tornberg} applied orthorhombic grids\cite{lindbo-tornberg}, and show that error estimates for gridding operations derived using orthorhombic grids are insufficient to bound the error on deformed grids. 
				
		To derive a bound for the quadrature error, it is easiest to consider the grid-to-particle operations. The exact particle velocity is computed by summing the fluid velocity weighted by the Gaussian quadrature kernel at every grid point,
		\begin{equation}
			{\bf U} = \int_{\Omega}\limits d{\bf y} \, \tilde{\bf H}( {\bf y}) \, \left( \frac{2\xi^{2}}{\pi\eta} \right)^{3/2} \, e^{-2\xi^{2}{\bf r}\cdot{\bf r}/\eta}, \label{eqn:quad1}
		\end{equation}
		where ${\bf r} = {\bf y} - {\bf x}$ is the minimum image distance from the center of a particle located at ${\bf x}$ to the lattice point located at ${\bf y}$, $\tilde{\bf H}( {\bf y} )$ represents the fluid velocity at $ \bf y $, and the subscript $\Omega$ denotes an integral over the entire system volume. We seek to bound the error in computing the trapezoidal rule quadrature approximation to this integral truncated over a finite number of grid points occupying a finite volume in the simulation space,
		\begin{equation}
			{\bf Q} = h^{3} \, \sum\limits_{{\bf n}\in P } \, \tilde{\bf H}\left({\bf n}h\right)\,K\left({\bf r}_{n}\right), \label{eqn:quadapprox}
		\end{equation}
		where ${\bf r}_{n} = {\bf n}h - {\bf x}$ is the distance from the grid point ${\bf n}$ to the particle center, $h^{3}$ is the volume of a grid element and $P$ is the number of grid points within the support of each particle in each direction. The set of grid points $\{{\bf n}\}$ contains all grid points within the support of the particle and the function $K$ represents the Gaussian kernel
		\begin{equation}
			K\left({\bf r}\right) = \left(\frac{2\xi^{2}}{\pi\eta}\right)^{3/2}e^{-2\xi^{2}{\bf r}\cdot{\bf r}/\eta}.
		\end{equation}
				
		The error associated with the quadrature approximation to the integral is
		\begin{align}
			\epsilon_{q} &= \left\lVert h^{3} \, \sum\limits_{{\bf n}\in P } \, \tilde{\bf H}\left({\bf n}h\right)\,K\left({\bf r}_{n}\right) - \int\limits_{\Omega} d{\bf y} \, \tilde{\bf H}\left({\bf y}\right) \, K\left({\bf r}\right) \right\rVert \\
						 &\leq \left\lVert h^{3} \, \sum\limits_{{\bf n}} \, \tilde{\bf H}\left({\bf n}h\right)\,K\left({\bf r}_{n}\right) - \int\limits_{\Omega} d{\bf y} \, \tilde{\bf H}\left({\bf y}\right) \, K\left({\bf r}\right) \right\rVert \label{eqn:A}\\
						 &\qquad + \left\lVert h^{3} \, \sum\limits_{{\bf n}\notin P} \, \tilde{\bf H}\left({\bf n}h\right)\,K\left({\bf r}_{n}\right) \right\rVert , \nonumber
		\end{align}
		where ${\bf r} = {\bf A} \cdot {\bf z}$ given that ${\bf z}$ is the distance from the center of the particle to the orthorhombic (not deformed) set of grid points, and deformation of the lattice vectors away from an orthorhombic orientation is assumed to be described by the volume-preserving linear deformation tensor ${\bf A}$. 			
		
\begin{figure}
			\includegraphics{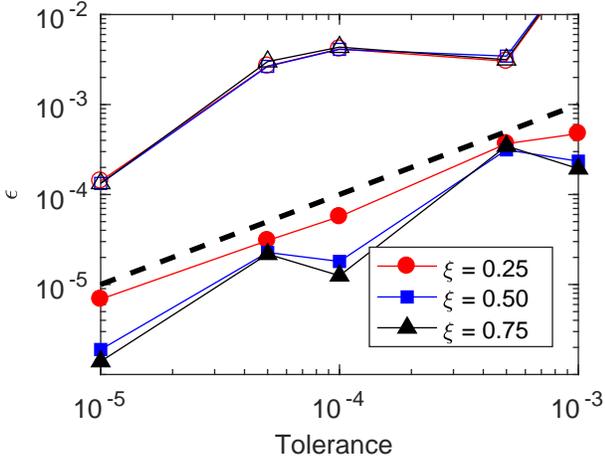}
			\caption{Relative error in the sedimentation velocity of a suspension of hard sphere particles on a sheared FFT lattice (total strain, $\gamma = 0.5$) as a function of the specified tolerance at various values of the splitting parameter, $\xi$. The error is defined as $\epsilon=\lVert {\bf U} - {\bf U}_{\rm exact} \rVert / \lVert {\bf U}_{\rm exact} \rVert$, where the exact solution was computed using an orthorhombic grid and an error tolerance of $10^{-10}$. The quadrature parameters are chosen using either the sheared grid bounds in Equation \eqref{eqn:bound} (closed symbols), or the bounds for orthorhombic grids from \citeauthor{lindbo-tornberg} (open symbols). The orthorhombic error bounds correspond to setting $\lambda_{\rm max} = 1$ in \eqref{eqn:bound}. The dashed line denotes equality between the computed error and the specified tolerance, $\epsilon={\rm tolerance}$.}
			\label{fig:shear}
		\end{figure}

		When appropriately normalized, $\lvert \tilde{\bf H} \rvert \leq 1$ over all space, so the estimate derived by setting $\tilde{\bf H} = 1$  everywhere provides a bound on the control of the approximation to the integral. The first term in Equation \eqref{eqn:A} can be bound starting with Poisson's summation formula, $h^{3}\sum_{\bf n} {\bf f}\left({\bf n}h\right)=\sum_{\boldsymbol{\kappa}}=\hat{\bf f}\left(\boldsymbol{\kappa}/h\right)$, where ${\bf f} = \tilde{\bf H}\,K$ and $\hat{\bf f}$ denotes the Fourier transform of ${\bf f}$. Using this formula, the term can be written as
		\begin{equation}
			\left\lVert \sum\limits_{\boldsymbol{\kappa}\neq{\bf 0}} \hat{\bf f}\left(\boldsymbol{\kappa}/h\right) \right\rVert \leq C_{A} \left\lVert \sum\limits_{\boldsymbol{\kappa}\neq{\bf 0}} \hat{K}\left(\boldsymbol{\kappa}/h\right) \right\rVert, \label{eqn:poisson}
		\end{equation}
		where the ${\bf k}={\bf 0}$ terms cancels with the integral in Equation \eqref{eqn:A}, $C_{A}$ is an unknown order-one constant, and the inequality holds because $\tilde{\bf H}$ is uniform in the system volume. The sum expressed in Equation \eqref{eqn:poisson} only contains terms which are exponentially decaying. Therefore, the value of the sum can be bounded. Substituting the expression for $\hat{K}$ and using the fact that the most slowly decaying term, $\lVert \boldsymbol{\kappa}\rVert=1$, dominates the sum gives the bound
		\begin{align}
			\left\lVert \sum\limits_{\boldsymbol{\kappa}\neq{\bf 0}} \hat{\bf f}\left(\boldsymbol{\kappa}/h\right) \right\rVert \leq C_{A}^{\prime}\frac{e^{-\pi^{2}\eta\,{\bf k}\cdot\left({\bf A}^{T}\cdot{\bf A}\right)^{-1}\cdot{\bf k}/2\xi^{2}h^{2}}}{{\rm det}\left({\bf A}^{T}\cdot{\bf A}\right)^{1/2}},
		\end{align}
		where $C_{A}^{\prime}$ is a different unknown order-one constant from $C_{A}$. A final simplication can be made by making the substitutions that $e^{-{\bf k}\cdot\left({\bf A}^{T}\cdot{\bf A}\right)^{-1}\cdot{\bf k}} \leq e^{-{\bf k}\cdot{\bf k}/\lambda_{\rm max}}$ and ${\rm det}\left({\bf A}\right)=1$. The determinant is unity because the deformation is volume-preserving and $\lambda_{\rm max}$ is the maximum eigenvalue of ${\bf A}^{T}\cdot{\bf A}$. Therefore, the bound on the first norm in \eqref{eqn:A} is
		\begin{equation}
			C_{A}^{\prime} \, e^{-\pi^{2}\eta/2\xi^{2}h^{2}\lambda_{\rm max}} \label{eqn:bound1}
		\end{equation}
		The second term in Equation \eqref{eqn:A} can be bounded by the integral of the Gaussian kernel over the region exterior to the particle's support because $\tilde{\bf H}$ is bounded over all space
		\begin{equation}
			\left\lVert h^{3} \, \sum\limits_{{\bf n}\notin P} \, \tilde{\bf H}\left({\bf n}h\right)\,K\left({\bf r}_{n}\right) \right\rVert \leq \left\lVert \, \int\limits_{{\bf z}\cdot{\bf A}^{T}\cdot{\bf A}\cdot{\bf z} \geq h^{2}P^{2}/4} d{\bf z} \, K\left({\bf z}\right) \right\rVert
		\end{equation}
		On a deformed grid, the region neglected during quadrature is the volume outside the ellipsoid defined by ${\bf z}\cdot{\bf A}^{T}\cdot{\bf A}\cdot{\bf z} = h^{2}P^{2}/4$. Integration is straightforward after a transformation from spherical coordinates ${\bf z}$ to ellipsoidal coordinates ${\bf r} = {\bf A} \cdot {\bf z}$ and yields
		\begin{equation}
			\frac{\lVert {\bf A}^{-1} \rVert}{{\rm det}\left({\bf A}^{T}\cdot{\bf A}\right)^{1/2}} \, {\rm erfc}\left[ \sqrt{\frac{2\xi^{2}}{\eta}} \, R \right] \label{eqn:bound2}
		\end{equation}
		where $R = \left[ \left( {\bf r} \cdot \left({\bf A}^{T}\cdot{\bf A}\right)^{-1} \cdot {\bf r} \right) |_{{\bf r}\cdot{\bf r}=h^{2}P^{2}/4} \right]^{1/2}$ and can be bounded by $ R \leq \left( \lambda_{\rm max} \, {\bf r}\cdot{\bf r} \right)^{1/2} = hP/2\lambda_{\rm max}^{1/2}$. Again, $\lambda_{\rm max}$ is the maximum eigenvalue of ${\bf A}^{T}\cdot{\bf A}$, and ${\bf A}$ represents volume-preserving deformations so ${\rm det}\left({\bf A}\right)=1$. Making these substitutions in \eqref{eqn:bound2}, combined with \eqref{eqn:bound1}, the error bound can expressed as
		\begin{equation}
			\epsilon_{q} \leq C \left( e^{-\pi^{2}\eta/2\xi^{2}h^{2}\lambda_{\rm max}} + {\rm erfc}\left[ \sqrt{\frac{2\xi^{2}}{\eta}} \, \frac{hP}{2\lambda_{\rm max}^{1/2}} \right] \right),
		\end{equation}
		where $C$ is an unknown order-one constant. The Gaussian kernels are usually expressed in terms of a resolution width $m$ and number of support nodes $P$, which are related to the decay constant $\eta= \left(hP\xi/m\right)^{2}$ as described by \citeauthor{lindbo-tornberg}\cite{lindbo-tornberg},
		\begin{equation}
			\epsilon_{q} \leq C \left[ e^{-\pi^{2}P^{2}/2m^{2}\lambda_{\rm max}} + {\rm erfc}\left( \frac{m}{\sqrt{2\lambda_{\rm max}}} \right) \right]. \label{eqn:bound}
		\end{equation}
		In the case that $\lambda_{\rm max}=1$, this expression is identical to the error bound for orthorhombic grids derived by \citeauthor{lindbo-tornberg}. Equation \eqref{eqn:bound} shows that the presence of deformation decreases the error decay rate by a factor of $\lambda_{\rm max}$, necessitating larger supports for constant accuracy. Figure \ref{fig:shear} shows the relative error in the sedimentation velocity of a random suspension of hard spheres with a simulation box that has undergone linear shear with a strain of $\gamma = 0.5$. Using the error bounds for an orthorhombic grid to choose the quadrature parameters for a deformed grid produces a calculation error that is an order of magnitude larger than the specified tolerance.
		
		Linear shear is described by the tensor
		\begin{equation}
			{\bf A}_{\rm shear} = \begin{bmatrix}
				1 & \gamma & 0 \\ 0 & 1 & 0 \\ 0 & 0 & 1
			\end{bmatrix},
		\end{equation}
		and the associated maximum eigenvalue required in the error bound is
		\begin{equation}
			\lambda_{\rm max}\left( {\bf A}_{\rm shear}^{T}\cdot{\bf A}_{\rm shear} \right) = 1 + \frac{\gamma^{2}}{2} + \gamma \, \sqrt{1 + \frac{\gamma^{2}}{4}} \label{eqn:lmax1}.
		\end{equation}
		For planar extensional flow, the deformation tensor is
		\begin{equation}
			{\bf A}_{\rm ext} = \begin{bmatrix}
				\varepsilon & 0 & 0 \\ 0 & 1/\varepsilon & 0 \\ 0 & 0 & 1
			\end{bmatrix},
		\end{equation}
		and the associated maximum eigenvalue is
		\begin{equation}
			\lambda_{\rm max}\left( {\bf A}_{\rm ext}^{T}\cdot{\bf A}_{\rm ext} \right) = \begin{cases} \varepsilon^{2} & , \varepsilon \geq 1 \\ 1/\varepsilon^{2} & , \varepsilon < 1 \end{cases} \label{eqn:lmax2}.
		\end{equation}
		Equations \eqref{eqn:lmax1} and \eqref{eqn:lmax2} are tabulated in Table \ref{tab:lmax} for various values of $\gamma$, which shows that even moderate strain values can produce large relative changes in $\lambda_{\rm max}$ and therefore strongly effect the value of $\epsilon_{q}$. For example, if simulation parameters were chosen to satisfy a tolerance of $\epsilon_{q} = 10^{-3}$ with $\gamma = 0.5$ ($\varepsilon=0.5$) using the bounds for orthorhombic lattice (i.e. setting $\lambda_{\rm max} = 1$ in Equation \eqref{eqn:bound}), the actual expected error would be $\epsilon_{q,{\rm actual}}^{\rm shear} \approx \left(10^{-3}\right)^{1/\lambda_{\rm max}^{\rm shear}} = 0.015$ for simple shear and $\epsilon_{q,{\rm actual}}^{\rm ext} \approx \left(10^{-3}\right)^{1/\lambda_{\rm max}^{\rm ext}} = 0.046$ for planar extension, both of which are more than an order of magnitude larger than the desired tolerance. 
		\begin{table}
			\caption{Maximum eigenvalue of ${\bf A}^{T}\cdot{\bf A}$ for different values of $\gamma$ for both simple shear and planar extensional lattice deformations. The Gaussian kernel support size for an error bound of $10^{-3}$ is also reported.}
			\begin{tabular}{ccc|ccc}
				\hline\hline
				$\gamma$ & $\lambda_{\rm max}^{\rm shear}$ & $P_{\rm shear}$ & $\varepsilon$ & $\lambda_{\rm max}^{\rm ext}$ & $P_{\rm ext}$ \\
				0.0 & 1.00 & 5 & 1.0 & 1.00 & 5 \\
				0.1 & 1.11 & 6 & 1.1 & 1.21 & 6 \\
				0.5 & 1.64 & 7 & 1.5 & 2.25 & 8 \\
				1.0 & 2.62 & 8 & 2.0 & 4.00 & 10 \\
				\hline\hline
			\end{tabular}
			\label{tab:lmax}
		\end{table}	
		To control the numerical error in the simulations, the two contributions to the error in Equation \eqref{eqn:bound} are set equal, and the complementary error function is bounded by ${\rm erfc}\left(x\right) \leq e^{-x^{2}}$ for $x>0$	which sets the value of $m$ relative to $P$ as $m = \sqrt{\pi P}$. Then the support size, $P$ is chosen using this relation and \eqref{eqn:bound} along with a desired tolerance. Increasing the strain, $\gamma$, increases the support size required to maintain the error at a given tolerance. The values of $P$ required to maintain an error of $\epsilon_{q}=10^{-3}$ for a range of $\gamma$ values is given in Table \ref{tab:lmax}.
	
\section{Fast Stochastic Sampling for Brownian Dynamics}
\label{sec:Sampling}
	The equation of motion for colloidal particles is derived from the definition of the particle velocity,
	\begin{align}
		\frac{d{\bf x}}{dt} = \mathcal{U}, \label{eqn:veldef}
	\end{align}
	and the overdamped force balance
	\begin{align}
		{\bf 0} = \mathcal{F}^{H} + \mathcal{F}^{B} + \mathcal{F}^{P}, \label{eqn:Fbal}
	\end{align}
	where $\mathcal{F}^{H}$ is the hydrodynamic force exerted by the fluid on the particle, $\mathcal{F}^{B}$ is the stochastic Brownian force exerted on the particle due to collisions with the solvent, and $\mathcal{F}^{P}$ represents any other conservative force acting on the particles, e.g. inter-particle attraction/repulsion. The hydrodynamic force is related to the particle velocity through the mobility relation,
	\begin{equation}
		\begin{bmatrix}
			\mathcal{U} \\ {\bf 0}
		\end{bmatrix} = -\mathcal{M} \cdot \begin{bmatrix}
			\mathcal{F}^{H} \\ {\bf S}^{H}
		\end{bmatrix}, \label{eqn:constraint}
	\end{equation}
	where the rate of strain is zero for rigid particles. Together equations \eqref{eqn:veldef}-\eqref{eqn:constraint} constitute a stochastic differential algebraic equation (SDAE) for the particle positions. This SDAE is index-1, which means that the algebraic constraints can be explicitly solved, reducing the system to a stochastic differential equation. 	
	
	The solution to equation \eqref{eqn:constraint} is the same as that to equation \eqref{eqn:constrained_vel},
	\begin{align}
		\mathcal{U} = -{\bf R}_{\mathcal{FU}}^{-1} \cdot \mathcal{F}^{H}, \label{eqn:constrained_vel2}
	\end{align}
	where ${\bf R}_{\rm FU} = \left( \mathcal{M}_{\mathcal{UF}} - \mathcal{M}_{\mathcal{U}{\rm S}} \cdot \mathcal{M}_{\rm ES}^{-1} \cdot \mathcal{M}_{{\rm E}{\mathcal F}} \right)^{-1}$ is the resistance tensor, which is the $\mathcal{FU}$ component of $\mathcal{M}^{-1}$,
	\begin{equation}
		\mathcal{M}^{-1} = \begin{bmatrix} {\bf R}_{\mathcal{FU}} & {\bf R}_{\mathcal{F}{\rm E}} \\ {\bf R}_{{\rm S}\mathcal{U}} & {\bf R}_{\rm SE} \end{bmatrix}
	\end{equation}	
	It should be noted that we do not have analytical expressions for the multiparticle resistance tensors in general, and ${\bf R}_{\mathcal{FU}}$ specifically, so in simulations operations with these tensor can only be evaluated with linear solves of the constrained system of equations \eqref{eqn:constraint}. 
	
	Equation \eqref{eqn:constrained_vel2} can be substituted for the hydrodynamic force in \eqref{eqn:Fbal}
	\begin{align}
		{\bf 0} = -{\bf R}_{\mathcal{FU}}\cdot\mathcal{U} + \mathcal{F}^{B} + \mathcal{F}^{P}.
	\end{align}
	This algebraic equation can be substituted into \eqref{eqn:veldef} to produce the differential equation governing particle motion
	\begin{align}
		\frac{d{\bf x}}{dt} = {\bf R}_{\mathcal{FU}}^{-1} \cdot \mathcal{F}^{P} + {\bf R}_{\mathcal{FU}}^{-1} \cdot \mathcal{F}^{B}. \label{eqn:eom}
	\end{align}		
	The fluctuation-dissipation theorem states that the Brownian force has zero mean and covariance proportional to the resistance tensor,
	\begin{align}
		\overline{\mathcal{F}^{B}\left(t\right)\,\mathcal{F}^{B}\left(t^{\prime}\right)} = 2k_{B}T \, \delta\left(t-t^{\prime}\right) \, {\bf R}_{\mathcal{FU}}.
	\end{align}
	The Brownian force that satisfies this distribution is sampled from a collection of standard independent Wiener processes, ${\bf W}(t)$
	\begin{align}
		\mathcal{F}^{B} = \sqrt{2k_{B}T} \, {\bf R}_{\mathcal{FU}}^{1/2} \cdot \frac{d{\bf W}}{dt}, \label{eqn:FB}
	\end{align}
	where ${\bf R}_{\mathcal{FU}}^{1/2}$ is the denotes the ``square root" of the resistance tensor and can be any matrix that satisfies ${\bf R}_{\mathcal{FU}}^{1/2}\cdot\left({\bf R}_{\mathcal{FU}}^{1/2}\right)^{\dagger} = {\bf R}_{\mathcal{FU}}$ where superscript dagger denotes the adjoint. Substituting the form for the Brownian force in \eqref{eqn:FB} into differential equation for particle motion \eqref{eqn:eom}, the change in particle position $d{\bf x}$ during the incremental time $dt$ is
	\begin{align}
		d{\bf x} = {\bf R}_{\mathcal{FU}}^{-1} \cdot \mathcal{F}^{P} \, dt + \sqrt{2k_{B}T} \, {\bf R}_{\mathcal{FU}}^{-1/2} \cdot d{\bf W}(t). \label{eqn:stratonovich}
	\end{align}
	In computer simulations, the solution to the stochastic differential equation \eqref{eqn:stratonovich} over a given time interval $t\in\left[t_{o},t_{f}\right]$ is approximated piecewise by the collection of solutions on $n$ sub-intervals of $\left[t_{o},t_{f}\right]$. In the limit that $n \rightarrow \infty$ the piecewise approximation to the solution converges to the differential equation \cite{wongzakai}
	\begin{align}
		d{\bf x} &= {\bf R}_{\mathcal{FU}}^{-1} \cdot \mathcal{F}^{P} \, dt + \sqrt{2k_{B}T} \, {\bf R}_{\mathcal{FU}}^{-1/2} \cdot d{\bf W}(t) + \label{eqn:ito}\\ &\qquad k_{B}T \boldsymbol{\nabla}\cdot{\bf R}_{\mathcal{FU}}^{-1} \, dt. \nonumber 
	\end{align}
	It is clear from equation \eqref{eqn:ito} that the divergence of the mobility, $k_{B}T \boldsymbol{\nabla}\cdot{\bf R}_{\mathcal{FU}}^{-1}$, also called the thermal drift term, which arises from the discretization of the differential equation must be calculated to capture the proper particle dynamics. Methods for computing the divergence of ${\bf R}_{\mathcal{FU}}^{-1}$ are discussed in section \ref{sec:Integrator}. 
	
	The PSE algorithm provides a rapid approach to generate the stochastic particle displacements with controlled accuracy. Generating fluctuations with covariance defined by ${\bf R}_{\mathcal{FU}}^{-1}$ is more challenging than when the covariance is defined by $\mathcal{M}_{\rm UF}$, as was the case in the unconstrained RPY level of hydrodynamic approximation. The difficulty arises because the resistance tensor ${\bf R}_{\mathcal{FU}}$ includes constrained dynamics, i.e. the linear solve of the stresslet, and therefore cannot be directly sampled using the PSE method. However, displacements consistent with equation \eqref{eqn:ito} can be generated by combining the PSE method with integration schemes for SDAEs, as will be described in section \ref{sec:Integrator}. 
	
	The fundamental idea behind the PSE method is to sample two independent stochastic displacements with different covariances, $\mathcal{M}_{r}$ and $\mathcal{M}_{w}$, such that the sum of the two has the same distribution as displacements sampled with covariance $\mathcal{M}$.\cite{pse} Mathematically, this is written as
	\begin{equation}
		\mathcal{M}^{1/2} \cdot d {\bf W}\left(t\right) 
                \,{\buildrel d \over =}\
                \mathcal{M}_{r}^{1/2} \cdot d {\bf W}_{1}\left(t\right) + \mathcal{M}_{w}^{1/2} \cdot d {\bf W}_{2}\left(t\right), \label{eqn:split}
	\end{equation}
	where ${\bf W}_{1}$ and ${\bf W}_{2}$ are two independent Wiener processes and ${\buildrel d \over =}$ means equality in distribution. The sampling operations associated with $\mathcal{M}_{r}^{1/2}$ and $\mathcal{M}_{w}^{1/2}$ are accelerated by leveraging the algebraic structures of the two different operators, as described in section \ref{sec:MR12} and \ref{sec:MW12}.
	
	\subsection{$\mathcal{M}$ is Fundamentally Ill-Conditioned}	
	\label{sec:CondM}
	Before describing the PSE sampling approach in detail, it is worth remarking on the necessity of such a split sampling technique for large scale simulations. Conventional approaches to BD simulations use iterative methods (Chebyshev polynomials, Krylov subspaces) to approximate the action of the square root of the mobility tensor on a vector\cite{fixman}, but the number of iterations grows rapidly with increasing condition number. This presents a challenge for performing rapid simulations because the mobility tensor is ill-conditioned for even modest numbers of particles, and its condition number increases with the number of particles modeled. 
	
	A simple argument for the system size dependence of the condition number can be constructed by considering a hydrodynamically coupled cluster of $N$ particles and size $L$. The condition number can be estimated by approximating the smallest eigenvalue $\lambda_{\rm min}$ and largest eigenvalue $\lambda_{\rm max}$ of the UF component of the mobility tensor for the cluster ($\mathcal{M}_{\rm UF}$). The smallest eigenvalue represents uncorrelated particle motion and is therefore described by the single particle mobility, $\lambda_{\rm min} \sim 1/\eta \, {a}$. The largest eigenvalue represents motion of all $N$ particles correlated across the cluster size $L$ and so depends on the number of particles and the decay rate of the interactions, $\lambda_{\rm max} \sim N/\eta \, {L}$. The cluster size can be related to the number of particles and its fractal dimension, $L{\sim}a\,N^{1/d_{f}}$, so $\lambda_{\rm max} \sim N^{(d_{f}-1)/d_{f}}/\eta \, {a}$. Therefore, the dependence of condition number of $\mathcal{M}_{\rm UF}$ on the size of the cluster is $\kappa\left(\mathcal{M}_{\rm UF}\right) \sim \lambda_{\rm max}/\lambda_{\rm min} \sim N^{(d_{f}-1)/d_{f}}$. In many physical systems, e.g. polymers, gels, and random hard spheres, the fractal dimension is between two and three, so the scaling of the condition number, $\kappa\left(\mathcal{M}_{\rm UF}\right) $, varies between $ O(N^{1/2}) $ to $ O(N^{2/3}) $ in these limits.
	
	Similar condition number estimates can be derived for other mobility sub-blocks of the mobility tensor, e.g. the torque-rotation $\mathcal{M}_{{\Omega}{\rm L}}$ and stresslet-strain $\mathcal{M}_{\rm ES}$ couplings. In these cases, where the ${\rm UF}$ interaction decays with the inverse distance between particles $r^{-1}$, the $\Omega{\rm L}$ and ${\rm ES}$ interactions decay with the inverse distance cubed $r^{-3}$. This does not affect the estimate of the smallest eigenvalue, but the largest eigenvalue now scales as $\lambda_{\rm max} \sim N/\eta \, {L}^{3} \sim N^{(d_{f}-3)/d_{f}}/\eta \, {a}$, so the condition number is $\kappa\left(\mathcal{M}_{\Omega{\rm L}}\right) \sim \kappa\left(\mathcal{M}_{\rm ES}\right) \sim N^{(d_{f}-3)/d_{f}}$. For fractal dimensions ranging from two to three, the condition number, $\kappa\left(\mathcal{M}_{\Omega{\rm L}}\right), \kappa\left(\mathcal{M}_{\rm ES}\right) $ scales between $ O( N^{-1/2} ) $ and $ O( N^{0} )$.
	
	It is apparent from these estimates that $\mathcal{M}$ is ill-conditioned because the ${\rm UF}$ interactions are fundamentally ill-conditioned. Because $\mathcal{M}_{\rm UF}$, and therefore $\mathcal{M}$, is increasingly ill-conditioned for larger numbers of particles, conventional simulation methods employing iterative methods for the square root of $\mathcal{M}$ are limited in the systems sizes they can attain, typically to $\mathcal{O}\left( 10^{3}-10^{4} \right)$ particles. New methods that circumvent the conditioning problem, such as the PSE method\cite{pse} and its extension here, are needed to perform large scale simulations. Circumventing the ill-conditioning of the mobility tensor increases the systems sizes accessible to simulations to $\mathcal{O}\left( 10^{5}-10^{7} \right)$ particles on current GPU hardware.

	\subsection{Real Space Square Root}
	\label{sec:MR12}
	The action of $\mathcal{M}_{r}^{1/2}$ on a vector can be efficiently approximated by the same iterative schemes that are applied to $\mathcal{M}^{1/2}$ for arbitrarily large numbers of particles because $\mathcal{M}_{r}$ contains only short-ranged (exponential decay) interactions. In particular, we use the Lanczos method proposed by \citeauthor{chow-saad}\cite{chow-saad} for its superior convergence and numerical error control. The exponentially decaying nature of the real-space scalar mobility functions for finite $\xi\,a > 0$ guarantees a compact eigenspectrum for $\mathcal{M}_{r}$. To show this, the condition number of $\mathcal{M}_{r}$ can be estimated in the same way as in section \ref{sec:CondM}. The smallest eigenvalue remains unchanged, while the largest eigenvalue depends only on the exponentially decaying interactions within a neighborhood of size $ \xi^{-1}$, $\lambda_{\rm max} \sim \phi \, \left( \xi \, a \right)^{-d_{f}} / \eta \, \xi^{-1} $, where $\phi$ is the local particle volume fraction. Therefore, for given value of $\xi$, the condition number of $\mathcal{M}_{r}$ is $\kappa\left(\mathcal{M}_{r}\right) \sim \phi \, \left( \xi \, a \right)^{1-d_{f}} $, which is constant and independent of system size. The value of $\kappa\left(\mathcal{M}_{r}\right)$ is also $\mathcal{O}\left(1\right)$ for the modest values of $\xi$ typically used in simulations ($0.1 \leq \xi \leq 1$ in most applications). This means that the efficiency of the square root calculation for $\mathcal{M}_{r}$ is solely determined by the choice of the Ewald splitting parameter, $\xi$.

	\subsection{Wave Space Square Root}
	\label{sec:MW12}
	The wave space component of the mobility can be written as a sequence of linear operations using the NUDFT as in equation \eqref{eqn:WSops}
	\begin{equation}
		\mathcal{M}_{w} = \mathbb{D}^{-1} \cdot {\bf P}^{\dagger} \cdot {\bf B} \cdot {\bf P} \cdot \mathbb{D},
	\end{equation}
	where the NUFFT can be used to evaluate the Fourier transform by making the substitution that $\mathbb{D} = \mathbb{F} \cdot {\bf Q}$.
	
	The square root of the wave space mobility can be expressed analytically by noting that the matrix ${\bf B}$ is block diagonal and positive semi-definite and therefore has an analytical square root. Using the unitary property of the Fourier transform allows $\mathcal{M}_{w}$ to be re-expressed as
	\begin{equation}
		\mathcal{M}_{w} = \left( \mathbb{D}^{\dagger} \cdot {\bf P}^{\dagger} \cdot {\bf B}^{1/2} \right) \cdot \left( \mathbb{D}^{\dagger} \cdot {\bf P}^{\dagger} \cdot {\bf B}^{1/2} \right)^{\dagger}. 
	\end{equation}
	This demonstrates that, as in the RPY case, the Brownian displacement can be calculated with a single matrix multiplication
	\begin{equation}
		\mathcal{M}_{w}^{1/2} \cdot d{\bf W}_{2}(t) = \mathbb{D}^{\dagger} \cdot {\bf P}^{\dagger} \cdot {\bf B}^{1/2} \cdot d{\bf W}_{2}(t). \label{eqn:DFTsqrt}
	\end{equation}
	The random variable $d{\bf W}_{2}$ is complex-valued Gaussian variate that generates real-valued velocity fields when acted upon by $\mathcal{M}_{w}^{1/2}$. Ensuring the real-valuedness of the velocity fields generated by $\mathcal{M}_{w}^{1/2} \cdot d{\bf W}_{2}$ requires enforcing the conjugacy condition $d{\bf W}_{2}\left(-{\bf k}\right)=d{\bf W}_{2}^{\dagger}\left({\bf k}\right)$. 
	
	Using the NUFFT, the square root is written as
	\begin{equation}
		\mathcal{M}_{w}^{1/2} = \sigma^{1/2}\,{\bf Q}^{T}\cdot\mathbb{F}^{\dagger}\cdot{\bf P}^{\dagger}\cdot{\bf B}^{1/2}, \label{eqn:FFTsqrt}
	\end{equation}
	which is applied to the vector of complex-valued Gaussian variates $d{\bf W}_{2}$ by a combination of rescaling (${\bf B}^{1/2}$), projection (${\bf P}^{\dagger}$), inverse fast Fourier transform ($\mathbb{F}^{\dagger}$), and interpolation/quadrature (${\bf Q}^{T}$). 
	
	Because all of the long-ranged (slowly decaying) contributions to the mobility sum are contained in the wave space part, sampling the wave space square root using the direct calculation in \eqref{eqn:DFTsqrt} or \eqref{eqn:FFTsqrt} entirely avoids expensive iterative methods which suffer from ill-conditioning. This most troublesome part of the noise calculation is handled in a single matrix multiplication. 
		
	\subsection{Integration Scheme for Constrained Dynamics}
	\label{sec:Integrator}
	The displacements sampled through the operations of $\mathcal{M}_{r}^{1/2}$ and $\mathcal{M}_{w}^{1/2}$ described in sections \ref{sec:MR12} and \ref{sec:MW12} are associated with unconstrained particle motion and do not account for thermal drift. Many computational integration schemes could be used to incorporate the constraints and the thermal drift into the displacements, and we will use a simple two-step scheme that only requires a single constraint solve \cite{fcm}. The scheme contains four steps:
	\begin{enumerate}
		\item Generate stochastic slip velocities at time step $k$ using the PSE method, ignoring the stresslet constraint
			\begin{align}
				\begin{bmatrix}	\mathcal{U}^{B}_{k} \\ {\bf E}^{B}_{k}	\end{bmatrix} &= \mathcal{M}_{r,k}^{1/2} \cdot d{\bf W}_{1}(t) + \mathcal{M}_{w,k}^{1/2} \cdot d{\bf W}_{2}(t)
			\end{align}
		\item Advance to the midpoint using the unconstrained slip velocities
			\begin{align}
				{\bf x}_{k+1/2} &= {\bf x}_{k} + \frac{\Delta t}{2} \, \mathcal{U}^{B}_{k}
			\end{align}
		\item Solve the constrained problem at the midpoint for the true particle velocity
			\begin{align}
				\begin{bmatrix}	\mathcal{U}_{k+1/2} \\ -{\bf E}^{B}_{k}	\end{bmatrix} &= \mathcal{M}_{k+1/2} \cdot \begin{bmatrix} {\bf 0} \\ {\bf S}_{k+1/2} \end{bmatrix} + \begin{bmatrix} \mathcal{U}^{B}_{k} \\ {\bf 0} \end{bmatrix}
			\end{align}
		\item Advance from the initial point to the final point using the constrained midpoint velocity
			\begin{align}
				{\bf x}_{k+1} &= {\bf x}_{k} + {\Delta t} \, \mathcal{U}_{k+1/2}
			\end{align}
	\end{enumerate}
	In the preceding equations, the subscript $k$ denotes the integration time point. The displacements generated by this scheme implicitly account for the divergence of the mobility, $\boldsymbol{\nabla}\cdot{\bf R}_{\mathcal{FU}}^{-1}$, and are consistent with Equation \eqref{eqn:ito} to $\mathcal{O}\left(\Delta t\right)$. A proof of the weak accuracy is shown in Appendix \ref{app:AppendixB}. 
	
	Other fast approximation schemes for the divergence of the mobility exist, for example random finite differences (RFD)\cite{fib}, which can be combined with a simple Euler-Maruyama integration scheme. The RFD algorithm requires three solves of the stresslet constraint per time step; two to compute the approximation to $\boldsymbol{\nabla}\cdot{\bf R}_{\mathcal{FU}}^{-1}$ and one to evaluate the deterministic and stochastic contributions to particle motion. The stresslet solve is the most expensive part of the calculation by an order of magnitude, so a midpoint integration scheme is expected to execute 67\% faster than a single step integration scheme employing the RFD drift algorithm.  
		
\section{Results and Discussion}
	The stresslet-constrained PSE method was implemented on graphics processing units (GPUs) using CUDA to leverage the massively parallel computing architecture. The linked-cell algorithm used to compute the real space sum leverages state-of-the-art cell list and neighbor list algorithms implemented in the HOOMD-blue software suite \cite{hoomd1,hoomd2}. The real space mobility function tabulations are stored on-device in textured memory for fast access by the GPU kernels. The FFTs required to evaluate the wave space Ewald sum are performed using the CUDA library CUFFT to optimize performance. The spreading operator ${\bf Q}$ uses particle-based atomic additions with a thread block assigned to each particle to transfor particle forces to the grid. A particle-based thread block algorithm with parallel reduction is used to rapidly evaluate the interpolation operator ${\bf Q}^{\dagger}$. 
	
	In this section we assess the accuracy and performance of the stresslet-constrained PSE algorithm. The accuracy is validated against known equilibrium and dynamic properties of particulate dispersions. We first demonstrate that the integration scheme for the Langevin equation \eqref{eqn:ito} reproduces the correct Boltzmann distribution for a pair of interacting particles. The multiparticle distributions for non-interacting and hard sphere particles are quantified by the structure factor for a range of particle volume fractions and match known results. We then demonstrate the accuracy of the stresslet caclulation by comparing the computed hydrodynamic function and short-time self-diffusivity to known results. The algorithm performance is measured for a wide range of system sizes $N=8000-512000$ and volume fractions $\phi=0.1-0.5$, and the optimal performance is shown to be independent of both quantities. Performance comparison with the original PSE algorithm for the RPY description of hydrodynamic interactions shows a modest decrease in simulation throughput caused by increased data handling requirements and the required stresslet solve. For all accuracy and performance tests, the prescribed relative error tolerance for all calculations is $\epsilon=10^{-3}$ unless otherwise noted. 
	
	\subsection{Thermodynamic Equilibrium}
\begin{figure}
		\includegraphics{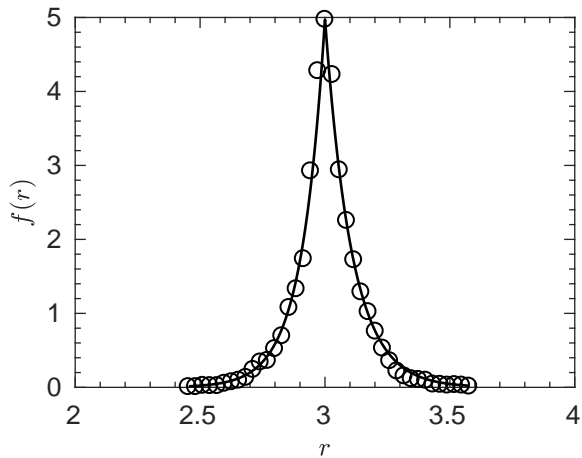}
		\caption{Probability distribution function for the distance between two particles connected by a constant spring. Circles represent simulation data and the solid line represents the analytical solution.}
		\label{fig:pairpdf}
	\end{figure}

	Capturing the correct equilibrium behavior in a stochastic simulation requires a consistent representation of the mobility in the stochastic displacement and the Brownian drift terms in the integration scheme for the particle equation of motion given in equation \eqref{eqn:ito}. A simple test of consistency in the integration scheme is to compute the distribution of center-to-center distances between a pair of freely diffusing particles connected by a prescribed interaction potential and compare the result to the analytically known Boltzmann distribution. In this test, the potential used is $U(r) = k \, \lvert r - r_{0} \rvert$, where $k$ is the spring constant, $r$ is the is magnitude of the center-to-center particle distance, and $r_{0}$ is the equilibrium particle separation. It can be shown that the Boltzmann distribution for the center-to-center distance is
	\begin{equation}
		f(r) = \frac{\tilde{k}^{3}}{2\left(2+\tilde{k}^{2}r_{0}^{2}\right)} \, r^{2} \, e^{ -\tilde{k} \, \lvert r - r_{0} \rvert},
	\end{equation}
	where $\tilde{k} = k/k_{B}T$. A comparison between this analytical result and the result for a simulation of two particles is shown in Figure \ref{fig:pairpdf}. Excellent agreement is observed between the simulation results and the analytical solution.
	
\begin{figure}
		\includegraphics{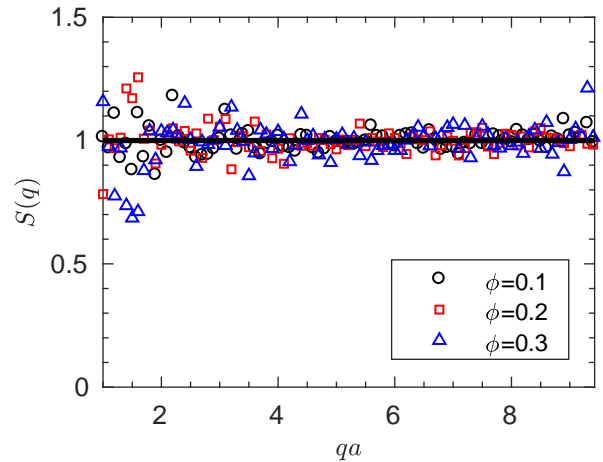}
		\caption{Static structure factor of 64000 non-interacting particles at varying volume fractions (symbols) compared with the theoretical result (line).}
		\label{fig:sofqig}
	\end{figure}

	The static structure factor provides a more rigorous test of the equilibrium particle distribution obtained in large scale simulations. The structure factor measures correlations between the positions of constituent elements in the observation volume. It is defined mathematically as proportional to the square magnitude of the Fourier transform of the particle number density,
	\begin{equation}
		S\left(q\right) = \frac{1}{N} \, \sum\limits_{j,k}^{N} e^{i \, {\bf q} \cdot \left( {\bf x}_{j} - {\bf x}_{k} \right) }. 
	\end{equation}
	For a randomly distributed set of coordinates, e.g. non-interacting particles, there are no positional correlations on average. Therefore, $S(q)$ is expected to be unity for all $q$. The structure factor computed from a dynamic simulation with stresslet constraints of 64000 non-interacting spherical particles is shown in Figure \ref{fig:sofqig} for a range of volume fractions. As expected, $S(q)=1$ for all $q$ and is independent of volume fraction. 
	
\begin{figure}
		\includegraphics{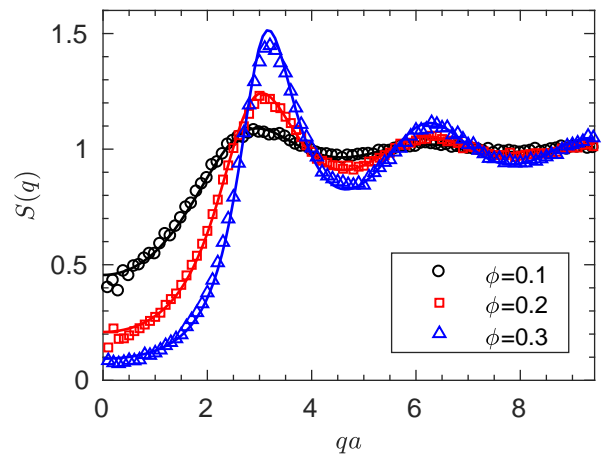}
		\caption{Static structure factor of 64000 hard sphere particles at varying volume fractions (circles) compared with the Percus-Yevick approximation (lines).}
		\label{fig:sofqhs}
	\end{figure}

	The same calculation can be made for particles interacting via hard sphere repulsions, for which there is an expected volume fraction and length scale dependence of the structure factor. The simulation results can be be compared to various theoretical results, for example the Percus-Yevick approximation. This comparison is shown in Figure \ref{fig:sofqhs} for simulations of 64000 hard sphere particles over a range of volume fractions. Excellent agreement between the simulation results and Percus-Yevick approximation is observed. 
	
	These three tests demonstrate that the algorithm and integration scheme reproduce the correct equilibrium behavior for a variety of scenarios. This indicates that the stochastic ${\bf R}_{\mathcal{FU}}^{-1/2}\cdot{d{\bf W}}$ and thermal drift $\boldsymbol{\nabla}\cdot{\bf R}_{\mathcal{FU}}^{-1}$ terms are represented consistently with one another and with the deterministic piece ${\bf R}_{\mathcal{FU}}^{-1}\cdot\mathcal{F}$.
	
	\subsection{Suspension Dynamics}
	Particle motions in dynamic simulation are a strong function of the level of hydrodynamic approximation used to represent the resistance tensor, $ \mathbf{R}_{\mathcal{FU}} $.  Higher accuracy approximations of the HI have a significant effect on the observed dynamics and transport properties of the simulated dispersion. Differences in the dynamics of simulations with and without the stresslet constraint, hereafter called RPY and constrained RPY, respectively, can be used to verify the accuracy of the constrained simulations. 
	
	One measure of the effect of HI on the particle dynamics is the hydrodynamic function, $H(q)$, which measures short-time wave-vector-dependent collective particle motions and can be computed from dynamic simulation using the short time decay of the dynamic structure factor
	\begin{equation}
		\frac{ S(q,t) }{ S(q,0) } = e^{-q^{2} \, D(q) \, t },
	\end{equation}
	where
	\begin{equation}
		H(q) = \frac{D(q)}{D_{0}} \, S(q)
	\end{equation}
	
\begin{figure}
		\includegraphics{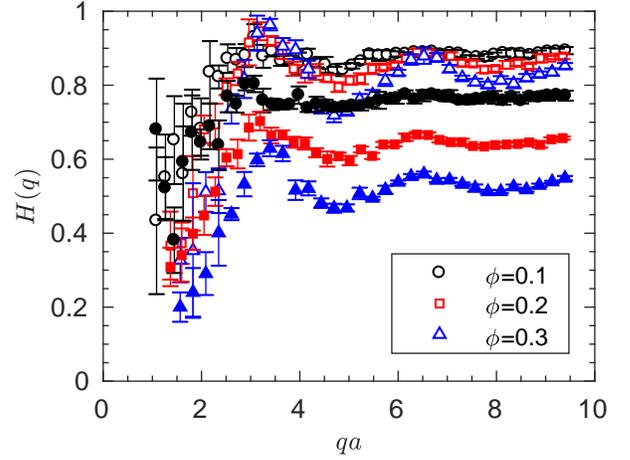}
		\caption{Hydrodynamic function of hard sphere particles at varying volume fractions calculated from simulations using RPY dynamics (open symbols) and stresslet-constrained RPY dynamics (filled symbols).}
		\label{fig:hofq}
	\end{figure}

	The hydrodynamic function for hard sphere particle suspensions computed from dynamic simulations using RPY and stresslet-constrained RPY representations of the HI is shown in Figure \ref{fig:hofq}. In the large length scale limit, corresponding to small wave vectors, $q \rightarrow 0$, $H(q)$ measures the mean velocity of collectively sedimenting particles. This value is expected to decrease with increasing volume fraction for both the RPY and constrained RPY simulations due to fluid backflow caused by volume conservation when a large collection of particles translates in unison\cite{hasimoto}. The volume fraction dependence of $H(q)$ is more pronounced in the constrained simulations because the stresslet constraint imposes additional viscous dissipation in the flow around the particles, which increases the hydrodynamic drag that they experience.
	
	 In the small length scale limit, $q \rightarrow \infty$, $H(q)$ measures the particle self-diffusion coefficient. The self-diffusion coefficient can also be computed from the trace of the $UF$ component of the mobility tensor\cite{phillips_PhysFluids_1988}. The self-diffusion coefficient, and therefore the large-$q$ limit of $H(q)$ is expected to be independent of volume fraction for the RPY simulations because the diagonal components of the RPY mobility tensor, $\mathcal{M}_{\rm UF}$, are constant and do not depend on the local suspension structure. In the constrained RPY simulations, the self-diffusivity is expected to decrease with increasing volume fraction because the diagonal components of the stresslet-constrained mobility tensor, ${\bf R}_{\mathcal{FU}}^{-1}$, depend on the local particle configuration and decrease with increasing volume fraction. The $H(q)$ data shown in Figure \ref{fig:hofq} is qualitatively consistent with the expected results across the full range of wave numbers. 
	
\begin{figure}
		\includegraphics{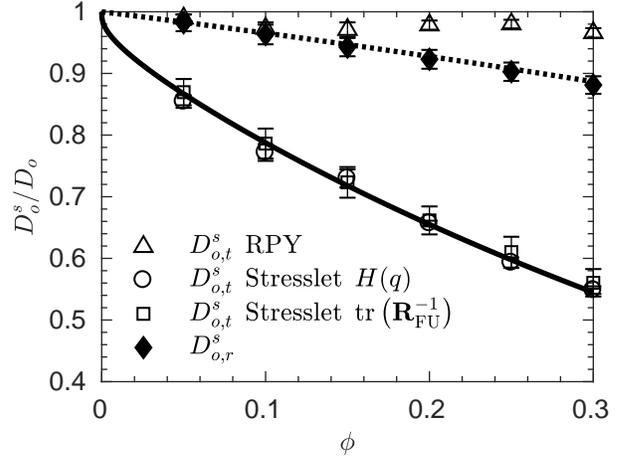}
		\caption{Translational (open symbols) and rotational (filled symbols) short-time self-diffusivity of hard sphere particles as a function of volume fraction. The translational $D_{o,t}^{s}$ is computed from the high-q limit of the hydrodynamic function for simulations with (circles) and without (triangles) the stresslet constraint, and from the trace of the stresslet-corrected mobility tensor (squares). The rotational $D_{o,r}^{s}$ is computed from the trace of the stresslet-corrected mobility tensor (filled diamonds). Simulation results using the mobility tensor with the stresslet constraint are compared with the equations of Banchio and Brady for the translational (solid line) and rotational (dotted line) self-diffusivity. }
		\label{fig:dssfromhofq}
	\end{figure}

	Although there is no analytical expression for $H(q)$, quantitative assessment of the $H(q)$ data can be made through the short-time self-diffusivity, $D_{o}^{s}$, which can be computed from either the large $q$ limit of $H(q)$ or from the trace of the mobility tensor. The translational and rotational short-time self-diffusivities are shown in Figure \ref{fig:dssfromhofq} and match very well to fits from \citeauthor{banchio}\cite{banchio}, indicating that the stresslet constraints are properly evaluated in the large scale dynamic simulations. 
	
	\subsection{Algorithm Performance}
\begin{figure}
		\includegraphics{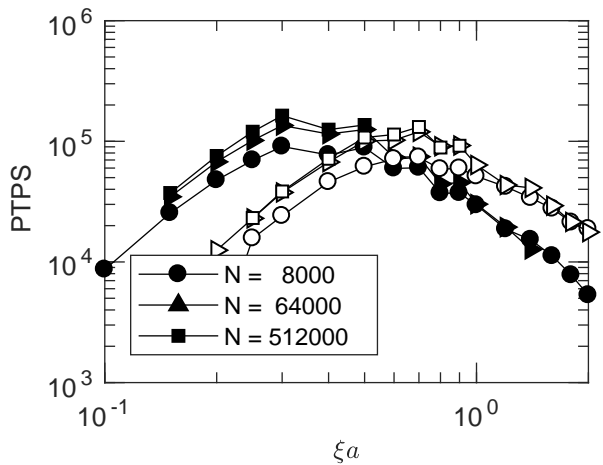}
		\caption{Timing results for calculations on random suspensions of hard spheres with volume fractions $\phi=0.1$ (closed symbols) and $\phi=0.5$ (open symbols) for different numbers of particles. The calculation cost of a full time step of the integration scheme, including the mobility calculation, stochastic sampling, and the stresslet solve is given in units of particle time steps per second (PTPS). }
		\label{fig:ptps}
	\end{figure}

\begin{figure}
		\includegraphics{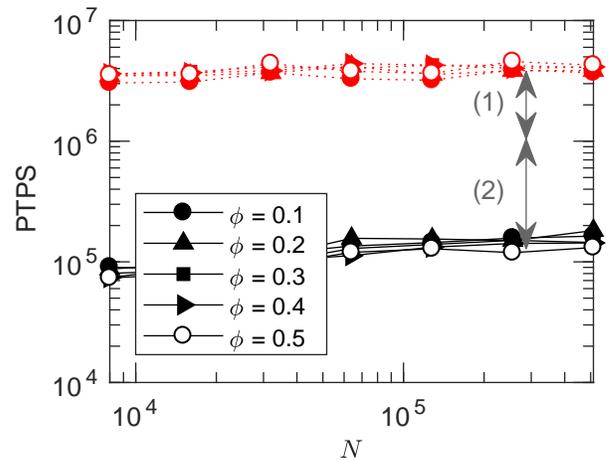}
		\caption{Optimal performance of the PSE (red dotted lines) and stresslet-constrained PSE (black solid lines) algorithms for calculations on random suspensions of hard spheres with volume fractions varying between $\phi=0.1$ and $\phi=0.5$ and system sizes varying from $N=8000$ to $N=512000$ particles. The performance decrease comes from two sources denoted in the Figure by grey arrows. Arrow (1) represents a factor of ${\approx}4$ increase in the amount of data that needs to be handled from 3 quantities per particle for the RPY case to 11 quantities per particle when the stresslet is included. Arrow (2) represents the $8-10$ iterations required to solve the stresslet constraint, which is independent of $\phi$ and $N$.}
		\label{fig:ptpsopt}
	\end{figure}

	The performance of the stresslet-constrained PSE algorithm was evaluated using an NVIDIA Tesla K40 GPU with the Maxwell architecture and a relative calculation error tolerance of $10^{-3}$ for all operations. Figure \ref{fig:ptps} is an example of the algorithm's performance for hard sphere configurations. The quantity reported is particle time steps per second (PTPS), which is the number of particles divided by the average time required to perform one time step worth of calculations. The algorithm performance is linear in the number of particles, $N$, and is optimizable with respect to the splitting parameter, $\xi$. The performance is expected to be independent of volume fraction for hard sphere dispersions, and is expected to increase as volume fraction decreases for dispersion with lower fractal dimension\cite{pse}. 
	Figure \ref{fig:ptpsopt} compares the optimal performance between the stresslet-constrained PSE algorithm and the original unconstrained RPY PSE algorithm for a range of volume fractions and system sizes. Including stresslets in the simulation reduces performance by a factor of $\approx 30$. There are 11 independent components of the force moments for each particle at the stresslet level of approximation: 3 for the force, 3 for the torque, and 5 for the stresslet, which is symmetric and traceless. The classic RPY case only has the 3 independent components of the force on each particle. This increased data volume reduces performance by a factor of $11/3$. Additionally, 8-10 iterations are needed to solve the stresslet constraints, compared to the single matrix multiplication required without constraints. Together, these two contributions account for the factor of 30 slow down when adding stresslets to the PSE algorithm. 
		
\section{Conclusion}
Our recently reported PSE algorithm for rapid Brownian dynamics simulations with HI\cite{pse} represented at the RPY level of approximation has been extended to include the torque and stresslet and the associated angular velocity and rate of strain. These higher order terms entered the mobility relation simply as an additional set of shape factors describing the velocity fields propagated by each force moment. As in the original PSE method, the Ewald sum of the mobility tensor (modified to include the torque and stresslet couplings) produced a positive splitting, i.e. the mobility tensor and both the real space and wave space contribution to its Ewald sum are positive semi-definite for all particle configurations. 

The stresslet constraint is particularly important in describing the dynamics of particle dispersions in linear flows. It is common and computationally convenient to represent these flows in simulations through deformed periodic lattices\cite{lees-edwards,kraynik-reinelt}. Green's function-based approaches to modeling the hydrodynamics, such as the method presented here, rely on the NUFFT to accelerate the periodic sums. The numerical error of such methods is well-controlled on orthorhombic lattices, but the orthorhombic error bounds do not extend to deformed grids. We have derived a bound on the NUFFT quadrature error on non-orthorhombic lattices and validated the bound against calculations of the numerical error. This error bound will be useful for all applications of the NUFFT on non-orthorhombic lattices. 

Including the stresslet, which is the first induced force moment in the multipole expansion, and determining its value required solving the linear equation constraining the local rate of strain on each (rigid) particle to zero. The proper formulation of the Langevin equation coupled with the mobility relation and the stresslet constraint is a stochastic differential algebraic equation (SDAE), where the differential variable is the particle position. Generating stochastic displacements consistent with the stresslet-constrained mobility tensor and integrating the SDAE were accomplished implicitly with a two step integration scheme. 

SDAEs can be used to represent a broad class of dynamic constraints, including the lubrication constraint in Stokesian Dynamics\cite{sd} and the rigid body constraint used when representing complex objects as rigid assemblies of spherical particles \cite{gang}. For each individual constraint, and for any combination of these constraints, the SDAE is index-1, which means that the algebraic variables can all be uniquely solved for and simple integration schemes can be applied. Much effort has recently been directed towards development and application of integration schemes for the Langevin equation describing hydrodynamically interacting particles, see for example \citeauthor{fib}\cite{fib} and \citeauthor{fcm}\cite{fcm}. Different schemes have been derived to handle various constraints and geometries, with effort directed at minimizing the number of constraint solves required. It is difficult to know \textit{a priori} which integration scheme is optimal for a given application, and how to accelerate the scheme. The SDAE framework disambiguates the choice of integration scheme by reducing problems to a common form. Future work will add the lubrication constraint to the stresslet formulation to complete development of a full Stokesian Dynamics implementation leveraging the PSE algorithm and SDAE representation. Because efficient methods exist to handle lubrication forces in simulations, Stokesian Dynamics using the PSE algorithm should be nearly as fast as the stresslet-constrained dynamic simulations in the present work. 

\section{Supplementary Material}
See supplementary material for for a GPU implementation of the PSE algorithm with stresslet constraints built as a plugin for HOOMD-blue. Included in the software is a sample script to perform a dynamic simulation using the plugin. 
	
\acknowledgements


J. Swan and A. Fiore gratefully acknowledge funding from the MIT Energy Initiative Shell Seed Fund and NSF Career Award CBET-1554398. Helpful conversations with Aleksandar Donev are gratefully acknowledged.


\begin{appendix}

\section{Real Space Scalar Mobility Functions}
\label{app:AppendixA}

	\subsection{Velocity-Force Mobility, ${\bf M}_{\rm UF}$}

	The scalar functions $F_{1}$ and $F_{2}$ are defined according to
	\begin{equation}
		{6\pi\eta{a}} \, M_{UF,im}^{\alpha\beta,(r)} = F_{1}(r) \, \left( \delta_{im} - \hat{r}_{i}\hat{r}_{m} \right) + F_{2}(r) \, \hat{r}_{i}\hat{r}_{m}
	\end{equation}
	where
	\begin{align}
		F_{1}(r) &= f_{0}^{(1)} + f_{1}^{(1)} \, e^{-r^2\xi^2} + f_{2}^{(1)} \, e^{-(r-2a)^2\xi^2} \\
		&\quad + f_{3}^{(1)} \, e^{-(r+2a)^2\xi^2} + f_{4}^{(1)}  \, {\rm erfc}\left(r\xi\right) \nonumber \\
		&\quad + f_{5}^{(1)} \, {\rm erfc}\left((r-2a)\xi\right) + f_{6}^{(1)} \, {\rm erfc}\left((r+2a)\xi\right) \nonumber \\
		F_{2}(r) &= f_{0}^{(2)} + f_{1}^{(2)} \, e^{-r^2\xi^2} + f_{2}^{(2)} \, e^{-(r-2a)^2\xi^2} \\
		&\quad + f_{3}^{(2)} \, e^{-(r+2a)^2\xi^2} + f_{4}^{(2)}  \, {\rm erfc}\left(r\xi\right) \nonumber \\
		&\quad + f_{5}^{(2)} \, {\rm erfc}\left((r-2a)\xi\right) + f_{6}^{(2)} \, {\rm erfc}\left((r+2a)\xi\right) \nonumber
	\end{align}
	\textbf{Case 1}, $r > 2a$
	
	First scalar mobility function:
	\begin{align*}
		f_{0}^{(1)} &= 0 \\
		f_{1}^{(1)} &= \frac{18 r^2 {\xi}^2+3}{64 \sqrt{\pi } a r^2 {\xi}^3} \\
		f_{2}^{(1)} &= \frac{2 {\xi}^2 (2 a-r) \left(4 a^2+4 a r+9 r^2\right)-2 a-3 r}{128 \sqrt{\pi } a r^3 {\xi}^3} \\
		f_{3}^{(1)} &= \frac{-2 {\xi}^2 (2 a+r) \left(4 a^2-4 a r+9 r^2\right)+2 a-3 r}{128 \sqrt{\pi } a r^3 {\xi}^3} \\
		f_{4}^{(1)} &= \frac{3-36 r^4 {\xi}^4}{128 a r^3 {\xi}^4} \\
		f_{5}^{(1)} &= \frac{4 {\xi}^4 (r-2 a)^2 \left(4 a^2+4 a r+9 r^2\right)-3}{256 a r^3 {\xi}^4} \\
		f_{6}^{(1)} &= \frac{4 {\xi}^4 (2 a+r)^2 \left(4 a^2-4 a r+9 r^2\right)-3}{256 a r^3 {\xi}^4}
	\end{align*}
	Second scalar mobility function:
	\begin{align*}
		f_{0}^{(2)} &= 0 \\
		f_{1}^{(2)} &= \frac{6 r^2 {\xi}^2-3}{32 \sqrt{\pi } a r^2 {\xi}^3} \\
		f_{2}^{(2)} &= \frac{-2 {\xi}^2 (r-2 a)^2 (2 a+3 r)+2 a+3 r}{64 \sqrt{\pi } a r^3 {\xi}^3} \\
		f_{3}^{(2)} &= \frac{2 {\xi}^2 (2 a+r)^2 (2 a-3 r)-2 a+3 r}{64 \sqrt{\pi } a r^3 {\xi}^3} \\
		f_{4}^{(2)} &= -\frac{3 \left(4 r^4 {\xi}^4+1\right)}{64 a r^3 {\xi}^4} \\
		f_{5}^{(2)} &= \frac{3-4 {\xi}^4 (2 a-r)^3 (2 a+3 r)}{128 a r^3 {\xi}^4} \\
		f_{6}^{(2)} &= \frac{3-4 {\xi}^4 (2 a-3 r) (2 a+r)^3}{128 a r^3 {\xi}^4}
	\end{align*}
	\textbf{Case 2}, $r \leq 2a$ 
	
	First scalar mobility function:
	\begin{align*}
		f_{0}^{(1)} &= -\frac{(r-2 a)^2 \left(4 a^2+4 a r+9 r^2\right)}{32 a r^3} \\
		f_{1}^{(1)} &= \frac{18 r^2 {\xi}^2+3}{64 \sqrt{\pi } a r^2 {\xi}^3} \\
		f_{2}^{(1)} &= \frac{2 {\xi}^2 (2 a-r) \left(4 a^2+4 a r+9 r^2\right)-2 a-3 r}{128 \sqrt{\pi } a r^3 {\xi}^3} \\
		f_{3}^{(1)} &= \frac{-2 {\xi}^2 (2 a+r) \left(4 a^2-4 a r+9 r^2\right)+2 a-3 r}{128 \sqrt{\pi } a r^3 {\xi}^3} \\
		f_{4}^{(1)} &= \frac{3-36 r^4 {\xi}^4}{128 a r^3 {\xi}^4} \\
		f_{5}^{(1)} &= \frac{4 {\xi}^4 (r-2 a)^2 \left(4 a^2+4 a r+9 r^2\right)-3}{256 a r^3 {\xi}^4} \\
		f_{6}^{(1)} &= \frac{4 {\xi}^4 (2 a+r)^2 \left(4 a^2-4 a r+9 r^2\right)-3}{256 a r^3 {\xi}^4} 
	\end{align*}
	Second scalar mobility function:
	\begin{align*}
		f_{0}^{(2)} &= \frac{(2 a-r)^3 (2 a+3 r)}{16 a r^3} \\
		f_{1}^{(2)} &= \frac{6 r^2 {\xi}^2-3}{32 \sqrt{\pi } a r^2 {\xi}^3} \\
		f_{2}^{(2)} &= \frac{-2 {\xi}^2 (r-2 a)^2 (2 a+3 r)+2 a+3 r}{64 \sqrt{\pi } a r^3 {\xi}^3} \\
		f_{3}^{(2)} &= \frac{2 {\xi}^2 (2 a+r)^2 (2 a-3 r)-2 a+3 r}{64 \sqrt{\pi } a r^3 {\xi}^3} \\
		f_{4}^{(2)} &= -\frac{3 \left(4 r^4 {\xi}^4+1\right)}{64 a r^3 {\xi}^4} \\
		f_{5}^{(2)} &= \frac{3-4 {\xi}^4 (2 a-r)^3 (2 a+3 r)}{128 a r^3 {\xi}^4} \\
		f_{6}^{(2)} &= \frac{3-4 {\xi}^4 (2 a-3 r) (2 a+r)^3}{128 a r^3 {\xi}^4}
	\end{align*}	
	\textbf{Case 3, Self Mobility} The real space part of the self mobility of a particle is given by the limit of equation \eqref{eqn:MUFr} as $r\rightarrow 0$, specifically $M_{UF,im}^{\alpha\alpha,(r)} = F_{1}(r{\rightarrow}0,\xi)\,\delta_{im}$:
	\begin{align}
		&{6\pi{\eta}a}\,M_{UF,im}^{\alpha\alpha,(r)} = \frac{ 1 - e^{-4a^{2}\xi^{2}} + 4 \pi^{1/2}a\xi \, {\rm erfc}\left(2a\xi\right)}{4 \pi^{1/2} \xi a} \,\delta_{im}. 
	\end{align}
	
	\subsection{Velocity-Couplet Mobility, ${\bf M}_{\rm UC}$}
	The scalar functions $G_{1}$ and $G_{2}$ are defined according to
	\begin{align}
		M_{UC,imn}^{\alpha\beta,(r)} &= G_{1} \, \left( \delta_{im}\hat{r}_{n} - \hat{r}_{i}\hat{r}_{m}\hat{r}_{m} \right) \\
									&\qquad + G_{2} \, \left( \delta_{in}\hat{r}_{m} + \delta_{mn}\hat{r}_{i} - 4\hat{r}_{i}\hat{r}_{m}\hat{r}_{n} \right) \nonumber 
	\end{align}
	where
	\begin{align}
		G_{1}(r) &= g_{0}^{(1)} + g_{1}^{(1)} \, e^{-r^2\xi^2} + g_{2}^{(1)} \, e^{-(r-2a)^2\xi^2}  \\
		&\quad + g_{3}^{(1)} \, e^{-(r+2a)^2\xi^2} + g_{4}^{(1)}  \, {\rm erfc}\left(r\xi\right) \nonumber \\
		&\quad + g_{5}^{(1)} \, {\rm erfc}\left((r-2a)\xi\right) + g_{6}^{(1)} \, {\rm erfc}\left((r+2a)\xi\right) \nonumber \\
		G_{2}(r) &= g_{0}^{(2)} + g_{1}^{(2)} \, e^{-r^2\xi^2} + g_{2}^{(2)} \, e^{-(r-2a)^2\xi^2} \\
		&\quad + g_{3}^{(2)} \, e^{-(r+2a)^2\xi^2} + g_{4}^{(2)}  \, {\rm erfc}\left(r\xi\right) \nonumber \\
		&\quad + g_{5}^{(2)} \, {\rm erfc}\left((r-2a)\xi\right) + g_{6}^{(2)} \, {\rm erfc}\left((r+2a)\xi\right) \nonumber
	\end{align}
	\textbf{Case 1}, $r > 2a$
	
	First scalar mobility function:
	\begin{widetext}
	\begin{align*}
		g_{0}^{(1)} &= 0 \\
		g_{1}^{(1)} &= \frac{-30 r^4 \xi^4+15 r^2 \xi^2+9}{64 \sqrt{\pi } a^4 r^3 \xi^5} \\
		g_{2}^{(1)} &= \frac{3 \xi^2 (2 a-r) \left(8 a^2+16 a r+25 r^2\right)-6 \xi^4 (2 a-r) \left(32 a^4+32 a^3 r+44 a^2 r^2+36 a r^3+25 r^4\right)-9 (2 a+5 r)}{640 \sqrt{\pi } a^4 r^4 \xi^5} \\
		g_{3}^{(1)} &= \frac{-3 \xi^2 (2 a+r) \left(8 a^2-16 a r+25 r^2\right)+6 \xi^4 (2 a+r) \left(32 a^4-32 a^3 r+44 a^2 r^2-36 a r^3+25 r^4\right)+18 a-45 r}{640 \sqrt{\pi } a^4 r^4 \xi^5} \\
		g_{4}^{(1)} &= \frac{60 r^6 \xi^6+9 r^2 \xi^2+9}{128 a^4 r^4 \xi^6} \\
		g_{5}^{(1)} &= -\frac{3 \left(512 a^6 \xi^6+5 r^2 \xi^2 \left(64 a^4 \xi^4+3\right)-256 a r^5 \xi^6+100 r^6 \xi^6+15\right)}{1280 a^4 r^4 \xi^6} \\
		g_{6}^{(1)} &= -\frac{3 \left(512 a^6 \xi^6+5 r^2 \xi^2 \left(64 a^4 \xi^4+3\right)+256 a r^5 \xi^6+100 r^6 \xi^6+15\right)}{1280 a^4 r^4 \xi^6}
	\end{align*}
	\end{widetext}
	Second scalar mobility function:
	\begin{widetext}
	\begin{align*}
		g_{0}^{(2)} &= 0 \\
		g_{1}^{(2)} &= \frac{-6 r^4 \xi^4+3 r^2 \xi^2-9}{64 \sqrt{\pi } a^4 r^3 \xi^5} \\
		g_{2}^{(2)} &= \frac{6 \xi^4 (r-2 a)^2 \left(16 a^3+24 a^2 r+14 a r^2+5 r^3\right)-3 \xi^2 \left(16 a^3+24 a^2 r+14 a r^2+5 r^3\right)+18 a+45 r}{640 \sqrt{\pi } a^4 r^4 \xi^5} \\
		g_{3}^{(2)} &= \frac{-6 \xi^4 (2 a+r)^2 \left(16 a^3-24 a^2 r+14 a r^2-5 r^3\right)+3 \xi^2 \left(16 a^3-24 a^2 r+14 a r^2-5 r^3\right)-18 a+45 r}{640 \sqrt{\pi } a^4 r^4 \xi^5} \\
		g_{4}^{(2)} &= \frac{12 r^6 \xi^6+9 r^2 \xi^2-9}{128 a^4 r^4 \xi^6} \\
		g_{5}^{(2)} &= \frac{12 \xi^6 \left(128 a^6-80 a^4 r^2+16 a r^5-5 r^6\right)-45 r^2 \xi^2+45}{1280 a^4 r^4 \xi^6} \\
		g_{6}^{(2)} &= \frac{12 \xi^6 \left(128 a^6-80 a^4 r^2-16 a r^5-5 r^6\right)-45 r^2 \xi^2+45}{1280 a^4 r^4 \xi^6}
	\end{align*}
	\end{widetext}
	\textbf{Case 2}, $r <= 2a$
	
	First scalar mobility function:
	\begin{widetext}
	\begin{align*}
		g_{0}^{(1)} &= -\frac{9 \left(r^2 \xi^2+1\right)}{128 a^4 r^4 \xi^6} \\
		g_{1}^{(1)} &= \frac{-30 r^4 \xi^4+15 r^2 \xi^2+9}{64 \sqrt{\pi } a^4 r^3 \xi^5} \\
		g_{2}^{(1)} &= \frac{3 \xi^2 (2 a-r) \left(8 a^2+16 a r+25 r^2\right)-6 \xi^4 (2 a-r) \left(32 a^4+32 a^3 r+44 a^2 r^2+36 a r^3+25 r^4\right)-9 (2 a+5 r)}{640 \sqrt{\pi } a^4 r^4 \xi^5} \\
		g_{3}^{(1)} &= \frac{-3 \xi^2 (2 a+r) \left(8 a^2-16 a r+25 r^2\right)+6 \xi^4 (2 a+r) \left(32 a^4-32 a^3 r+44 a^2 r^2-36 a r^3+25 r^4\right)+18 a-45 r}{640 \sqrt{\pi } a^4 r^4 \xi^5} \\
		g_{4}^{(1)} &= \frac{60 r^6 \xi^6+9 r^2 \xi^2+9}{128 a^4 r^4 \xi^6} \\
		g_{5}^{(1)} &= 0 \\
		g_{6}^{(1)} &= -\frac{3 \left(512 a^6 \xi^6+5 r^2 \xi^2 \left(64 a^4 \xi^4+3\right)+256 a r^5 \xi^6+100 r^6 \xi^6+15\right)}{1280 a^4 r^4 \xi^6}
	\end{align*}
	\end{widetext}
	Second scalar mobility function:
	\begin{widetext}
	\begin{align*}
		g_{0}^{(2)} &= -\frac{3 \left(128 a^6-80 a^4 r^2+16 a r^5-5 r^6\right)}{160 a^4 r^4} \\
		g_{1}^{(2)} &= \frac{-6 r^4 \xi^4+3 r^2 \xi^2-9}{64 \sqrt{\pi } a^4 r^3 \xi^5} \\
		g_{2}^{(2)} &= \frac{6 \xi^4 (r-2 a)^2 \left(16 a^3+24 a^2 r+14 a r^2+5 r^3\right)-3 \xi^2 \left(16 a^3+24 a^2 r+14 a r^2+5 r^3\right)+18 a+45 r}{640 \sqrt{\pi } a^4 r^4 \xi^5} \\
		g_{3}^{(2)} &= \frac{-6 \xi^4 (2 a+r)^2 \left(16 a^3-24 a^2 r+14 a r^2-5 r^3\right)+3 \xi^2 \left(16 a^3-24 a^2 r+14 a r^2-5 r^3\right)-18 a+45 r}{640 \sqrt{\pi } a^4 r^4 \xi^5} \\
		g_{4}^{(2)} &= \frac{12 r^6 \xi^6+9 r^2 \xi^2-9}{128 a^4 r^4 \xi^6} \\
		g_{5}^{(2)} &= \frac{12 \xi^6 \left(128 a^6-80 a^4 r^2+16 a r^5-5 r^6\right)-45 r^2 \xi^2+45}{1280 a^4 r^4 \xi^6} \\
		g_{6}^{(2)} &= \frac{12 \xi^6 \left(128 a^6-80 a^4 r^2-16 a r^5-5 r^6\right)-45 r^2 \xi^2+45}{1280 a^4 r^4 \xi^6}
	\end{align*}	
	\end{widetext}
		
	\subsection{Velocity Gradient-Couplet Mobility, ${\bf M}_{\rm DC}$}
	The scalar functions $K_{1}$, $K_{2}$, and $K_{3}$ are defined according to
	\begin{align}
		6\pi\eta{a}\,M_{DC,ijmn}^{\alpha\beta,(r)} &= K_{1} \left( \delta_{ij}\delta_{mn} + \delta_{im}\delta_{jn} -4 \delta_{in}\delta_{jm} \right) \\
										&\qquad + K_{2}\left( \delta_{jm}\hat{r}_{i}\hat{r}_{n} - \hat{r}_{i}\hat{r}_{j}\hat{r}_{m}\hat{r}_{n} \right) \nonumber \\
				  						&\qquad + K_{3} \left( \delta_{ij}\hat{r}_{m}\hat{r}_{n} + \delta_{im}\hat{r}_{j}\hat{r}_{n} + \delta_{jn}\hat{r}_{i}\hat{r}_{m} \right. \nonumber \\
				  						&\qquad \left. + \delta_{mn}\hat{r}_{i}\hat{r}_{j} + \delta_{in}\hat{r}_{j}\hat{r}_{m} - 6\hat{r}_{i}\hat{r}_{j}\hat{r}_{m}\hat{r}_{n} \right. \nonumber \\
				  						&\qquad \left. - \delta_{in}\delta_{jm} \right) \nonumber
	\end{align}
	where
	\begin{align}
		K_{1}(r) &= k_{0}^{(1)} + k_{1}^{(1)} \, e^{-r^2\xi^2} + k_{2}^{(1)} \, e^{-(r-2a)^2\xi^2}  \\
		&\quad + k_{3}^{(1)} \, e^{-(r+2a)^2\xi^2} + k_{4}^{(1)}  \, {\rm erfc}\left(r\xi\right) \nonumber \\
		&\quad + k_{5}^{(1)} \, {\rm erfc}\left((r-2a)\xi\right) + k_{6}^{(1)} \, {\rm erfc}\left((r+2a)\xi\right) \nonumber
	\end{align}
	\begin{align}
		K_{2}(r) &= k_{0}^{(2)} + k_{1}^{(2)} \, e^{-r^2\xi^2} + k_{2}^{(2)} \, e^{-(r-2a)^2\xi^2} \\
		&\quad + k_{3}^{(2)} \, e^{-(r+2a)^2\xi^2} + k_{4}^{(2)}  \, {\rm erfc}\left(r\xi\right) \nonumber \\
		&\quad + k_{5}^{(2)} \, {\rm erfc}\left((r-2a)\xi\right) + k_{6}^{(2)} \, {\rm erfc}\left((r+2a)\xi\right) \nonumber
	\end{align}
	\begin{align}
		K_{3}(r) &= k_{0}^{(3)} + k_{1}^{(3)} \, e^{-r^2\xi^2} + k_{2}^{(3)} \, e^{-(r-2a)^2\xi^2} \\
		&\quad + k_{3}^{(3)} \, e^{-(r+2a)^2\xi^2} + k_{4}^{(3)}  \, {\rm erfc}\left(r\xi\right) \nonumber \\
		&\quad + k_{5}^{(3)} \, {\rm erfc}\left((r-2a)\xi\right) + k_{6}^{(3)} \, {\rm erfc}\left((r+2a)\xi\right) \nonumber
	\end{align}
	\textbf{Case 1}, $r > 2a$
	
	First scalar mobility function:
	\begin{widetext}
	\begin{align*}
		k_{0}^{(1)} &= 0 \\
		k_{1}^{(1)} &= -\frac{3 \left(96 a^2 \xi^2 \left(2 r^4 \xi^4-r^2 \xi^2+3\right)-8 r^6 \xi^6+4 r^4 \xi^4+30 r^2 \xi^2-27\right)}{4096 \sqrt{\pi } a^6 r^4 \xi^7} \\
		k_{2}^{(1)} &= \frac{3 \left(6 \xi^2 (2 a+5 r) \left(12 a^2+5 r^2\right)-4 \xi^4 \left(96 a^5+144 a^4 r+64 a^3 r^2-30 a r^4-5 r^5\right)\right)}{40960 \sqrt{\pi } a^6 r^5 \xi^7} \\
			&\qquad + \frac{3 \left(8 \xi^6 (2 a-r)^3 \left(48 a^4+96 a^3 r+80 a^2 r^2+40 a r^3+5 r^4\right)+270 a-135 r\right)}{40960 \sqrt{\pi } a^6 r^5 \xi^7} \\
		k_{3}^{(1)} &= \frac{3 \left(-6 \xi^2 (2 a-5 r) \left(12 a^2+5 r^2\right)+4 \xi^4 \left(96 a^5-144 a^4 r+64 a^3 r^2-30 a r^4+5 r^5\right)\right)}{40960 \sqrt{\pi } a^6 r^5 \xi^7} \\
			&\qquad - \frac{3 \left(8 \xi^6 (2 a+r)^3 \left(48 a^4-96 a^3 r+80 a^2 r^2-40 a r^3+5 r^4\right)-135 (2 a+r)\right)}{40960 \sqrt{\pi } a^6 r^5 \xi^7} \\
		k_{4}^{(1)} &= \frac{24 \xi^2 \left(12 a^2 \left(4 r^6 \xi^6+3 r^2 \xi^2-3\right)-2 r^8 \xi^6+9 r^4 \xi^2-6 r^2\right)+81}{8192 a^6 r^5 \xi^8} \\
		k_{5}^{(1)} &= \frac{-1080 r^2 \xi^4 \left(4 a^2+r^2\right)+720 \xi^2 \left(6 a^2+r^2\right)+48 \xi^8 (r-2 a)^4 \left(48 a^4+96 a^3 r+80 a^2 r^2+40 a r^3+5 r^4\right)-405}{81920 a^6 r^5 \xi^8} \\
		k_{6}^{(1)} &= \frac{-1080 r^2 \xi^4 \left(4 a^2+r^2\right)+720 \xi^2 \left(6 a^2+r^2\right)+48 \xi^8 (2 a+r)^4 \left(48 a^4-96 a^3 r+80 a^2 r^2-40 a r^3+5 r^4\right)-405}{81920 a^6 r^5 \xi^8}
	\end{align*}
	\end{widetext}
	Second scalar mobility function:
	\begin{widetext}
	\begin{align*}
		k_{0}^{(2)} &= 0 \\
		k_{1}^{(2)} &= \frac{9 \left(32 a^2 \xi^2 \left(10 r^4 \xi^4+19 r^2 \xi^2+15\right)-56 r^6 \xi^6+28 r^4 \xi^4-78 r^2 \xi^2-45\right)}{4096 \sqrt{\pi } a^6 r^4 \xi^7} \\
		k_{2}^{(2)} &= \frac{9 \left(-8 \xi^6 (2 a-r) (2 a+r)^2 \left(16 a^4+16 a^2 r^2+7 r^4\right)-6 \xi^2 \left(8 a^3+20 a^2 r+46 a r^2-13 r^3\right)\right)}{8192 \sqrt{\pi } a^6 r^5 \xi^7} \\
			&\qquad + \frac{9 \left(4 \xi^4 (2 a-r) \left(16 a^4+32 a^3 r+48 a^2 r^2+56 a r^3+7 r^4\right)+45 (r-2 a)\right)}{8192 \sqrt{\pi } a^6 r^5 \xi^7} \\
		k_{3}^{(2)} &= \frac{9 \left(8 \xi^6 (r-2 a)^2 (2 a+r) \left(16 a^4+16 a^2 r^2+7 r^4\right)+6 \xi^2 \left(8 a^3-20 a^2 r+46 a r^2+13 r^3\right)\right)}{8192 \sqrt{\pi } a^6 r^5 \xi^7} \\
			&\qquad -\frac{9 \left(4 \xi^4 (2 a+r) \left(16 a^4-32 a^3 r+48 a^2 r^2-56 a r^3+7 r^4\right)+45 (2 a+r)\right)}{8192 \sqrt{\pi } a^6 r^5 \xi^7} \\
		k_{4}^{(2)} &= \frac{9 \left(8 \xi^2 \left(a^2 \left(-80 r^6 \xi^6+36 r^2 \xi^2+60\right)+r^2 \left(14 r^6 \xi^6+9 r^2 \xi^2-6\right)\right)-45\right)}{8192 a^6 r^5 \xi^8} \\
		k_{5}^{(2)} &= \frac{9 \left(8 \xi^2 \left(-9 r^2 \xi^2 \left(4 a^2+r^2\right)-60 a^2-2 \xi^6 \left(256 a^8+128 a^6 r^2-40 a^2 r^6+7 r^8\right)+6 r^2\right)+45\right)}{16384 a^6 r^5 \xi^8} \\
		k_{6}^{(2)} &= \frac{9 \left(8 \xi^2 \left(-9 r^2 \xi^2 \left(4 a^2+r^2\right)-60 a^2-2 \xi^6 \left(256 a^8+128 a^6 r^2-40 a^2 r^6+7 r^8\right)+6 r^2\right)+45\right)}{16384 a^6 r^5 \xi^8}
	\end{align*}
	\end{widetext}
	Third scalar mobility function:
	\begin{widetext}	
	\begin{align*}
		k_{0}^{(3)} &= 0 \\
		k_{1}^{(3)} &= \frac{9 \left(32 a^2 \xi^2 \left(-2 r^4 \xi^4+r^2 \xi^2+15\right)+8 r^6 \xi^6-4 r^4 \xi^4+18 r^2 \xi^2-45\right)}{4096 \sqrt{\pi } a^6 r^4 \xi^7} \\
		k_{2}^{(3)} &= \frac{9 \left(-8 \xi^6 (r-2 a)^2 (2 a+r)^3 \left(4 a^2+r^2\right)+4 \xi^4 (2 a+r)^3 \left(4 a^2+r^2\right)-6 \xi^2 (2 a+r)^2 (2 a+3 r)+45 (r-2 a)\right)}{8192 \sqrt{\pi } a^6 r^5 \xi^7} \\
		k_{3}^{(3)} &= \frac{9 \left(-8 \xi^6 (r-2 a)^3 (2 a+r)^2 \left(4 a^2+r^2\right)+4 \xi^4 (r-2 a)^3 \left(4 a^2+r^2\right)+6 \xi^2 (r-2 a)^2 (2 a-3 r)+45 (2 a+r)\right)}{8192 \sqrt{\pi } a^6 r^5 \xi^7} \\
		k_{4}^{(3)} &= \frac{9 \left(8 \xi^2 \left(4 a^2 \left(4 r^6 \xi^6-9 r^2 \xi^2+15\right)-2 r^8 \xi^6-3 r^4 \xi^2+6 r^2\right)-45\right)}{8192 a^6 r^5 \xi^8} \\
		k_{5}^{(3)} &= \frac{9 \left(16 \xi^8 \left(r^2-4 a^2\right)^3 \left(4 a^2+r^2\right)+24 r^2 \xi^4 \left(12 a^2+r^2\right)-48 \xi^2 \left(10 a^2+r^2\right)+45\right)}{16384 a^6 r^5 \xi^8} \\
		k_{6}^{(3)} &= \frac{9 \left(16 \xi^8 \left(r^2-4 a^2\right)^3 \left(4 a^2+r^2\right)+24 r^2 \xi^4 \left(12 a^2+r^2\right)-48 \xi^2 \left(10 a^2+r^2\right)+45\right)}{16384 a^6 r^5 \xi^8}
	\end{align*}
	\end{widetext}
	\textbf{Case 2}, $r <= 2a$
	
	First scalar mobility function:
	\begin{widetext}	
	\begin{align*}
		k_{0}^{(1)} &= -\frac{3 \left(768 a^8-640 a^6 r^2+256 a^3 r^5-120 a^2 r^6+5 r^8\right)}{2560 a^6 r^5} \\
		k_{1}^{(1)} &= -\frac{3 \left(96 a^2 \xi^2 \left(2 r^4 \xi^4-r^2 \xi^2+3\right)-8 r^6 \xi^6+4 r^4 \xi^4+30 r^2 \xi^2-27\right)}{4096 \sqrt{\pi } a^6 r^4 \xi^7} \\
		k_{2}^{(1)} &= \frac{3 \left(6 \xi^2 (2 a+5 r) \left(12 a^2+5 r^2\right)-4 \xi^4 \left(96 a^5+144 a^4 r+64 a^3 r^2-30 a r^4-5 r^5\right)\right)}{40960 \sqrt{\pi } a^6 r^5 \xi^7} \\
			&\qquad +\frac{3 \left(8 \xi^6 (2 a-r)^3 \left(48 a^4+96 a^3 r+80 a^2 r^2+40 a r^3+5 r^4\right)+270 a-135 r\right)}{40960 \sqrt{\pi } a^6 r^5 \xi^7} \\
		k_{3}^{(1)} &= \frac{3 \left(-6 \xi^2 (2 a-5 r) \left(12 a^2+5 r^2\right)+4 \xi^4 \left(96 a^5-144 a^4 r+64 a^3 r^2-30 a r^4+5 r^5\right)\right)}{40960 \sqrt{\pi } a^6 r^5 \xi^7} \\
			&\qquad -\frac{3 \left(8 \xi^6 (2 a+r)^3 \left(48 a^4-96 a^3 r+80 a^2 r^2-40 a r^3+5 r^4\right)-135 (2 a+r)\right)}{40960 \sqrt{\pi } a^6 r^5 \xi^7} \\
		k_{4}^{(1)} &= \frac{24 \xi^2 \left(12 a^2 \left(4 r^6 \xi^6+3 r^2 \xi^2-3\right)-2 r^8 \xi^6+9 r^4 \xi^2-6 r^2\right)+81}{8192 a^6 r^5 \xi^8} \\
		k_{5}^{(1)} &= \frac{-1080 r^2 \xi^4 \left(4 a^2+r^2\right)+720 \xi^2 \left(6 a^2+r^2\right)+48 \xi^8 (r-2 a)^4 \left(48 a^4+96 a^3 r+80 a^2 r^2+40 a r^3+5 r^4\right)-405}{81920 a^6 r^5 \xi^8} \\
		k_{6}^{(1)} &= \frac{-1080 r^2 \xi^4 \left(4 a^2+r^2\right)+720 \xi^2 \left(6 a^2+r^2\right)+48 \xi^8 (2 a+r)^4 \left(48 a^4-96 a^3 r+80 a^2 r^2-40 a r^3+5 r^4\right)-405}{81920 a^6 r^5 \xi^8}
	\end{align*}
	\end{widetext}
	Second scalar mobility function:
	\begin{widetext}
	\begin{align*}
		k_{0}^{(2)} &= \frac{3 \left(768 a^8 \xi^6-128 a^6 r^2 \xi^6-256 a^3 r^5 \xi^6+168 a^2 r^6 \xi^6-72 a^2 r^2 \xi^2-11 r^8 \xi^6+12 r^2\right)}{512 a^6 r^5 \xi^6} \\
		k_{1}^{(2)} &= \frac{9 \left(32 a^2 \xi^2 \left(10 r^4 \xi^4+19 r^2 \xi^2+15\right)-56 r^6 \xi^6+28 r^4 \xi^4-78 r^2 \xi^2-45\right)}{4096 \sqrt{\pi } a^6 r^4 \xi^7} \\
		k_{2}^{(2)} &= \frac{9 \left(-8 \xi^6 (2 a-r) (2 a+r)^2 \left(16 a^4+16 a^2 r^2+7 r^4\right)-6 \xi^2 \left(8 a^3+20 a^2 r+46 a r^2-13 r^3\right)\right)}{8192 \sqrt{\pi } a^6 r^5 \xi^7} \\
			&\qquad +\frac{9 \left(4 \xi^4 (2 a-r) \left(16 a^4+32 a^3 r+48 a^2 r^2+56 a r^3+7 r^4\right)+45 (r-2 a)\right)}{8192 \sqrt{\pi } a^6 r^5 \xi^7} \\
		k_{3}^{(2)} &= \frac{9 \left(8 \xi^6 (r-2 a)^2 (2 a+r) \left(16 a^4+16 a^2 r^2+7 r^4\right)+6 \xi^2 \left(8 a^3-20 a^2 r+46 a r^2+13 r^3\right)\right)}{8192 \sqrt{\pi } a^6 r^5 \xi^7} \\
			&\qquad -\frac{9 \left(4 \xi^4 (2 a+r) \left(16 a^4-32 a^3 r+48 a^2 r^2-56 a r^3+7 r^4\right)+45 (2 a+r)\right)}{8192 \sqrt{\pi } a^6 r^5 \xi^7} \\
		k_{4}^{(2)} &= \frac{9 \left(8 \xi^2 \left(a^2 \left(-80 r^6 \xi^6+36 r^2 \xi^2+60\right)+r^2 \left(14 r^6 \xi^6+9 r^2 \xi^2-6\right)\right)-45\right)}{8192 a^6 r^5 \xi^8} \\
		k_{5}^{(2)} &= \frac{24 \xi^2 \left(9 r^2 \xi^2 \left(4 a^2-3 r^2\right)-6 \left(30 a^2+r^2\right)+2 \xi^6 \left(-768 a^8+128 a^6 r^2+256 a^3 r^5-168 a^2 r^6+11 r^8\right)\right)+405}{16384 a^6 r^5 \xi^8} \\
		k_{6}^{(2)} &= \frac{9 \left(8 \xi^2 \left(-9 r^2 \xi^2 \left(4 a^2+r^2\right)-60 a^2-2 \xi^6 \left(256 a^8+128 a^6 r^2-40 a^2 r^6+7 r^8\right)+6 r^2\right)+45\right)}{16384 a^6 r^5 \xi^8}
	\end{align*}	
	\end{widetext}
	Third scalar mobility function:
	\begin{widetext}
	\begin{align*}
		k_{0}^{(3)} &= \frac{9 \left(4 a^2-r^2\right)^3 \left(4 a^2+r^2\right)}{512 a^6 r^5} \\
		k_{1}^{(3)} &= \frac{9 \left(32 a^2 \xi^2 \left(-2 r^4 \xi^4+r^2 \xi^2+15\right)+8 r^6 \xi^6-4 r^4 \xi^4+18 r^2 \xi^2-45\right)}{4096 \sqrt{\pi } a^6 r^4 \xi^7} \\
		k_{2}^{(3)} &= \frac{9 \left(-8 \xi^6 (r-2 a)^2 (2 a+r)^3 \left(4 a^2+r^2\right)+4 \xi^4 (2 a+r)^3 \left(4 a^2+r^2\right)-6 \xi^2 (2 a+r)^2 (2 a+3 r)+45 (r-2 a)\right)}{8192 \sqrt{\pi } a^6 r^5 \xi^7} \\
		k_{3}^{(3)} &= \frac{9 \left(-8 \xi^6 (r-2 a)^3 (2 a+r)^2 \left(4 a^2+r^2\right)+4 \xi^4 (r-2 a)^3 \left(4 a^2+r^2\right)+6 \xi^2 (r-2 a)^2 (2 a-3 r)+45 (2 a+r)\right)}{8192 \sqrt{\pi } a^6 r^5 \xi^7} \\
		k_{4}^{(3)} &= \frac{9 \left(8 \xi^2 \left(4 a^2 \left(4 r^6 \xi^6-9 r^2 \xi^2+15\right)-2 r^8 \xi^6-3 r^4 \xi^2+6 r^2\right)-45\right)}{8192 a^6 r^5 \xi^8} \\
		k_{5}^{(3)} &= \frac{9 \left(16 \xi^8 \left(r^2-4 a^2\right)^3 \left(4 a^2+r^2\right)+24 r^2 \xi^4 \left(12 a^2+r^2\right)-48 \xi^2 \left(10 a^2+r^2\right)+45\right)}{16384 a^6 r^5 \xi^8} \\
		k_{6}^{(3)} &= \frac{9 \left(16 \xi^8 \left(r^2-4 a^2\right)^3 \left(4 a^2+r^2\right)+24 r^2 \xi^4 \left(12 a^2+r^2\right)-48 \xi^2 \left(10 a^2+r^2\right)+45\right)}{16384 a^6 r^5 \xi^8}
	\end{align*}
	\end{widetext}
	\textbf{Case 3, Self Mobility} The real space part of the self mobility of a particle is given by the limit of equation \eqref{eqn:MDCr} as $r\rightarrow 0$, specifically $M_{DC,ijmn}^{\alpha\alpha,(r)} = K_{1}(r{\rightarrow}0,\xi)\,\left( \delta_{ij}\delta_{mn} + \delta_{im}\delta_{jn} -4 \delta_{in}\delta_{jm} \right)$:
	\begin{widetext}
	\begin{align}
		&{6\pi{\eta}a}\,M_{DC,ijmn}^{\alpha\alpha,(r)} = \left( -\frac{3 \left(6 a^2 \xi^2+1\right)}{80 \sqrt{\pi } a^6 \xi^3} + \frac{3  \left(10 a^2 \xi^2+1\right)}{80 \sqrt{\pi } a^6 \xi^3}\,e^{-4 a^2 \xi^2}-\frac{3 }{10 a^3}\,\text{erfc}(2 a \xi) \right) \,\left( \delta_{ij}\delta_{mn} + \delta_{im}\delta_{jn} -4 \delta_{in}\delta_{jm} \right). 
	\end{align}
	\end{widetext}

\section{Weak Accuracy of the Midpoint Integration Scheme}
\label{app:AppendixB}

The proposed integration scheme drifts particles from the initial point to the midpoint with the un-constrained Brownian slip velocity, then computes the actual velocity by solving the constrained problem at the midpoint,
\begin{align}
	{\bf x}_{k+1/2} &= {\bf x}_{k} + \frac{\Delta t}{2} \, \mathcal{U}^{B}_{k} \label{eqn:slip}\\
	\begin{bmatrix}	\mathcal{U}_{k+1/2} \\ -{\bf E}^{B}_{k}	\end{bmatrix} &= \mathcal{M}_{k+1/2} \cdot \begin{bmatrix} {\bf 0} \\ {\bf S}_{k+1/2} \end{bmatrix} + \begin{bmatrix} \mathcal{U}^{B}_{k} \\ {\bf 0} \end{bmatrix} \label{eqn:solve} \\
	{\bf x}_{k+1} &= {\bf x}_{k} + {\Delta t} \, \mathcal{U}_{k+1/2} \label{eqn:step}
\end{align}
where ${\bf U}^{B}$ and ${\bf E}^{B}$ have zero mean and covariance
\begin{equation}
	\left\langle \begin{bmatrix} \mathcal{U}^{B} \\ {\bf E}^{B} \end{bmatrix} \, \begin{bmatrix} \mathcal{U}^{B} \\ {\bf E}^{B} \end{bmatrix} \right\rangle = \frac{2k_{B}T}{\Delta t} \, \mathcal{M}.
\end{equation}
and the solution to \eqref{eqn:solve} for the velocity is
\begin{equation}
	\mathcal{U}_{k+1/2} = \mathcal{U}_{k}^{B} - \mathcal{M}_{\rm US}^{k+1/2} \cdot \left(\mathcal{M}_{\rm ES}^{k+1/2}\right)^{-1} \cdot {\bf E}_{k}^{B}. \label{eqn:midvel}
\end{equation}
A Taylor expansion of \eqref{eqn:midvel} about the initial point gives (changing to Einstein summation notation and dropping the subscript $k$ for the time step)
\begin{align}
	\mathcal{U}_{k+1/2,i} &\approx \mathcal{U}_{i}^{B} - \mathcal{M}_{{\rm US},ij} \left(\mathcal{M}_{\rm ES}\right)^{-1}_{jk} \, {E}_{k}^{B} + \\
		&\qquad \frac{\Delta t}{2} \, \mathcal{U}_{l}^{B} \, \nabla_{l} \, \left( \mathcal{U}_{i}^{B} - \mathcal{M}_{{\rm US},ij} \, \left(\mathcal{M}_{\rm ES}\right)^{-1}_{jk} \, E_{k}^{B} \right). \nonumber
\end{align}
The Brownian velocities are independent of the positional derivative in the Taylor expansion so
\begin{align}
	\mathcal{U}_{k+1/2,i} &\approx \mathcal{U}_{i}^{B} - \mathcal{M}_{{\rm US},ij} \left(\mathcal{M}_{\rm ES}\right)^{-1}_{jk} \, {E}_{k}^{B} - \\
		&\qquad \frac{\Delta t}{2} \, \mathcal{U}_{l}^{B} \, \nabla_{l} \, \left( \mathcal{M}_{{\rm US},ij} \, \left(\mathcal{M}_{\rm ES}\right)^{-1}_{jk} \right) \, E_{k}^{B} . \nonumber
\end{align}

\textbf{Displacement Mean}
The mean velocity is
\begin{align}
	&\langle \mathcal{U}_{k+1/2,i} \rangle \approx -\frac{\Delta t}{2} \, \left\langle \mathcal{U}_{l}^{B} \, \nabla_{l} \, \left( \mathcal{M}_{{\rm US},ij} \, \left(\mathcal{M}_{\rm ES}^{-1}\right)_{jk} \right) \, E_{k}^{B} \right\rangle \\
		&\qquad \approx -\frac{\Delta t}{2} \, \nabla_{l} \, \left( \mathcal{M}_{{\rm US},ij} \, \left(\mathcal{M}_{\rm ES}^{-1}\right)_{jk} \right) \, \left\langle E_{k}^{B} \mathcal{U}_{l}^{B} \right\rangle \\
		&\qquad \approx -k_{B}T \, \nabla_{l} \, \left( \mathcal{M}_{{\rm US},ij} \, \left(\mathcal{M}_{\rm ES}^{-1}\right)_{jk} \right) \, \mathcal{M}_{{\rm EF},kl} \\
		&\qquad \approx -k_{B}T \, \nabla_{i} \, \mathcal{M}_{{\rm US},ij} \, \left(\mathcal{M}_{\rm ES}^{-1}\right)_{jk}\,\mathcal{M}_{{\rm EF},kl} 
\end{align}
so the mean displacement, $\boldsymbol{\Delta}{\bf x} \equiv {\bf x}_{k+1}-{\bf x}_{k} = \Delta t \, \mathcal{U}_{k+1/2}$ is
\begin{equation}
	\left\langle \boldsymbol{\Delta}{\bf x} \right\rangle = k_{B}T{\Delta t} \, \grad \cdot {\bf R}_{\rm FU}^{-1} + \mathcal{O}\left({\Delta t}^{2}\right)
\end{equation}
because $ \grad \cdot \mathcal{M}_{\rm UF} = {\bf 0}$.

\textbf{Displacement Covariance}
The covariance of the velocity, again using the fact the the Brownian slip velocities are independent of position, is
\begin{widetext}
\begin{align}
	\left\langle \mathcal{U}_{k+1/2} \,\mathcal{U}_{k+1/2} \right\rangle &\approx \left\langle \mathcal{U}_{k}^{B}\,\mathcal{U}_{k}^{B} \right\rangle - 2\,\mathcal{M}_{\rm US}^{k} \cdot \left(\mathcal{M}_{\rm ES}^{k}\right)^{-1} \cdot \left\langle {\bf E}_{k}^{B}\mathcal{U}_{k}^{B} \right\rangle + \mathcal{M}_{\rm US}^{k} \cdot \left(\mathcal{M}_{\rm ES}^{k}\right)^{-1} \cdot \left\langle {\bf E}_{k}^{B} \, {\bf E}_{k}^{B} \right\rangle \cdot \left(\mathcal{M}_{\rm ES}^{k}\right)^{-1} \cdot \mathcal{M}_{\rm EF}^{k} \\
		&\approx \frac{2k_{B}T}{\Delta t}\,\left( \mathcal{M}_{\rm UF}^{k} - 2\,\mathcal{M}_{\rm US}^{k}\cdot\left(\mathcal{M}_{\rm ES}^{k}\right)^{-1} \cdot\mathcal{M}_{\rm EF}^{-1} + \mathcal{M}_{\rm US}^{k}\cdot\left(\mathcal{M}_{\rm ES}^{k}\right)^{-1} \cdot\mathcal{M}_{\rm EF}^{k} \right) \\
		&\approx \frac{2k_{B}T}{\Delta t}\,\left( \mathcal{M}_{\rm UF}^{k} - \,\mathcal{M}_{\rm US}^{k}\cdot\left(\mathcal{M}_{\rm ES}^{k}\right)^{-1} \cdot\mathcal{M}_{\rm EF}^{-1} \right)
\end{align}
\end{widetext}
So the covariance of the displacement is
\begin{align}
	\left\langle \boldsymbol{\Delta}{\bf x} \, \boldsymbol{\Delta}{\bf x} \right\rangle &= \left\langle {\Delta t}^{2} \mathcal{U}_{k+1/2}\,\mathcal{U}_{k+1/2} \right\rangle \\
		&= 2k_{B}T{\Delta t}\,{\bf R}_{\rm FU}^{-1} + \mathcal{O}\left({\Delta t}^{2}\right)
\end{align}

\section{Proof of Positive-Definiteness}
\label{app:AppendixC}
Here we present the proof for the positive-definiteness of the mobility tensor, extending the approach used in the original PSE work on the RPY tensor\cite{pse}, which followed the proof presented by \citeauthor{cichocki}\cite{cichocki} for the positive-definiteness for a tensor defined by the quadratic form
\begin{equation}
	\langle {\bf g} \lvert {\bf J} \rvert {\bf g} \rangle = \int d{\bf x} \, \int d{\bf y} \, {\bf g}^{\ast}\left({\bf x}\right) \cdot {\bf J}\left( {\bf x}, {\bf y} \right) \cdot {\bf g}\left( {\bf y} \right),
\end{equation}
where an asterisk denotes the complex conjugate and the integrals in ${\bf x}$ and ${\bf y}$ are over all space. The vectors ${\bf g}$ can be defined as
\begin{equation}
	{\bf g}\left( {\bf x} \right) \equiv \sum\limits_{i} \left[ {\bf w}^{(1)}_{i}\left({\bf x}\right) \:\: {\bf w}^{(2)}_{i}\left({\bf x}\right) \right] \cdot \left[ {\bf d}_{i}^{(1)} \:\: {\bf d}_{i}^{(2)} \right],
\end{equation}
where ${\bf w}_{i}^{(1)}$ and ${\bf w}_{i}^{(2)}$ are tensors such that ${\bf w}_{i}^{(1)}\cdot\mathcal{F}_{i}$ and ${\bf w}_{i}^{(2)}\cdot{\bf S}_{i}$ give the force density on the surface of particle $i$ caused by a generalized force (force and torque) and stresslet, respectively, and ${\bf d}_{i}^{(1)}$ and ${\bf d}_{i}^{(2)}$ are arbitrary vectors. The notation for ${\bf g}$ can be written more compactly as
\begin{equation}
	{\bf g}\left( {\bf x} \right) \equiv \sum\limits_{i} {\bf w}_{i} \cdot {\bf d}_{i},
\end{equation}
where ${\bf w}_{i} = \left[ {\bf w}^{(1)}_{i}\left({\bf x}\right) \:\: {\bf w}^{(2)}_{i}\left({\bf x}\right) \right]$ and ${\bf d}_{i} = \left[ {\bf d}^{(1)}_{i} \:\: {\bf d}^{(2)}_{i} \right]$. With these definitions for ${\bf w}_{i}$ and ${\bf d}_{i}$, the proof proceeds exactly as in the original work on the PSE algorithm \cite{pse}.

The pair mobility tensor is
\begin{equation}
	{\bf M}_{ij} = \left\langle {\bf w}_{i} \lvert {\bf J} \rvert {\bf w}_{i} \right\rangle.
\end{equation}
Because the Green's function ${\bf J}$ is positive semi-definite, and its quadratic form can be related to ${\bf M}_{ij}$, ${\bf M}_{ij}$ can be shown to be positive semi-definite as well:
\begin{equation}
	{\bf 0} \leq \left\langle {\bf g} \lvert {\bf J} \rvert {\bf g} \right\rangle = \sum\limits_{i,j} \, {\bf d}_{i}^{\ast} \cdot {\bf M}_{ij} \cdot {\bf d}_{j}
\end{equation}

The wave space representation of ${\bf J}$ is ${\bf J} = \sum_{{\bf k}\neq{\bf 0}} {\bf J}_{\bf k}$, where $0 \leq \langle {\bf g} \, \lvert \, {\bf J}_{\bf k} \, \rvert \, {\bf g} \rangle$. In this representation, the mobility is
	\begin{equation}
		{\bf M}_{ij} = \langle {\bf w}_{i} \, \lvert \, \sum\limits_{{\bf k}\neq{\bf 0}}{\bf J}_{\bf k} \, \rvert \, {\bf w}_{j} \rangle = \sum\limits_{{\bf k}\neq{\bf 0}} \, \langle {\bf w}_{i} \, \lvert \, {\bf J}_{\bf k} \, \rvert \, {\bf w}_{j} \rangle,
	\end{equation}	
	which can be decomposed into separate sums with the splitting function $H(k)$ to yield
	\begin{equation}
		{\bf M}_{ij} = \sum_{{\bf k}\neq{\bf 0}} \, \langle {\bf w}_{i} \, \lvert \, \left(1-H(k)\right)\,{\bf J}_{\bf k} \, \rvert \, {\bf w}_{j} \rangle + \sum_{{\bf k}\neq{\bf 0}} \, \langle {\bf w}_{i} \, \lvert \,  H(k)\,{\bf J}_{\bf k} \, \rvert \, {\bf w}_{j} \rangle,
	\end{equation}
	where the first term is the real space component of the mobility and the second term is the wave space component. It follows from $0 \leq H \leq 1$ and $0 \leq \langle {\bf g} \, \lvert \, {\bf J}_{\bf k} \, \rvert \, {\bf g} \rangle$ that
	\begin{align}
		0 \leq \sum_{{\bf k}\neq{\bf 0}} \langle {\bf g} \, \lvert \, (1-H(k)) \, {\bf J}_{\bf k} \, \rvert \, {\bf g} \rangle =  \sum_{i,k} {\bf d}_{i}^{\ast} \cdot {\bf M}^{(r)}_{ij} \cdot {\bf d}_{j}\\
		0 \leq \sum_{{\bf k}\neq{\bf 0}} \langle {\bf g} \, \lvert \, H(k) \, {\bf J}_{\bf k} \, \rvert \, {\bf g} \rangle = \sum_{i,j} {\bf d}_{i}^{\ast} \cdot {\bf M}^{(w)}_{ij} \cdot {\bf d}_{j}
	\end{align}
	which completes the proof that ${\bf M}^{(r)}_{ij}$ and ${\bf M}^{(w)}_{ij}$ are independently positive semi-definite. 

\end{appendix}
%

\end{document}